\documentclass[aps,prl,reprint,amsmath,amssymb,svgnames,dvipsnames,floatfix,superscriptaddress,nobibnotes]{revtex4-2}

\usepackage{graphicx}
\usepackage{dcolumn}
\usepackage{bm}
\usepackage{physics}
\usepackage{pgfplots}
\usepgfplotslibrary{groupplots}
\pgfplotsset{compat=newest}

\usepackage{algorithm}
\usepackage{silence}
\usepackage{algpseudocode}

\usepackage[caption=false]{subfig}
\usepackage[version=4]{mhchem}

\setcitestyle{super}

\begin{document}

\title{Advancing Natural Orbital Functional Calculations Through Deep Learning-Inspired Techniques for Large-Scale Strongly Correlated Electron Systems}

\author{Juan Felipe Huan Lew-Yee}
\email{felipe.lew.yee@dipc.org}
\affiliation{Donostia International Physics Center (DIPC), 20018 Donostia, Spain.}

\author{Jorge M. del Campo}
\email{jmdelc@unam.mx}
\affiliation{Departamento de F\'isica y Qu\'imica Te\'orica, Facultad de Qu\'imica,
Universidad Nacional Aut\'onoma de M\'exico, M\'exico City, C.P. 04510,
M\'exico}

\author{Mario Piris}
\email{mario.piris@ehu.eus}
\affiliation{Donostia International Physics Center (DIPC), 20018 Donostia, Spain.}
\affiliation{Kimika Fakultatea, Euskal Herriko Unibertsitatea (UPV/EHU), 20018 Donostia, Spain.}
\affiliation{IKERBASQUE, Basque Foundation for Science, 48013 Bilbao, Spain.}

\date{\today}

\begin{abstract}
Natural orbital functional (NOF) theory offers a promising approach for studying strongly correlated systems at an affordable computational cost, with an accuracy comparable to highly demanding wavefunction-based methods. However, its widespread adoption in cases involving a large number of correlated electrons has been limited by the extensive iterations required for convergence. In this work, we present a disruptive approach that embeds the techniques used for optimization in deep learning within the NOF calculation, constituting a substantial advance in the scale of accessible systems. The revamped procedure is based on the adaptive momentum technique for orbital optimization, alternated with the optimization of the occupation numbers, significantly improving the computational feasibility of challenging calculations. This work represents a complete change in the size scale of the systems that can be reached using NOF theory. We demonstrate this with three examples that involve a large number of electrons: (i) the symmetric dissociation of a large hydrogen cluster, (ii) an analysis of occupancies distribution in fullerenes, and (iii) a study of the singlet-triplet energy gap in linear acenes. Notably, the hydrogen cluster calculation, featuring 1000 electrons, represents the largest NOF calculation performed to date and one of the largest strongly correlated electron calculations ever reported. This system, which serves as an ideal model for a strongly correlated Mott insulator, illustrates a metal-to-insulator transition where all electrons participate in the correlation phenomenon, offering insight in a unique challenge. We anticipate that this work will enable the practical application of NOFs to increasingly complex and intriguing systems, leveraging the method's inherent scalability and accuracy.
\end{abstract}

\maketitle

Studying strongly correlated systems often requires computationally expensive wavefunction-based methods, which become intractable as the number of correlated electrons increases. This challenge has sparked significant interest in developing more cost-effective methods that can deliver accurate results. In this context, natural orbital functional (NOF) theory emerges as a promising alternative for electronic structure calculations, particularly in cases where standard density functional approximations lack accuracy and capturing strong correlation effects becomes crucial.

The NOF is the energy functional of the one-particle reduced density matrix ($\mathbf{\Gamma}$) in the natural spin-orbital representation,\cite{Lowdin1955a} where it takes a diagonal form: 
\begin{equation} 
\Gamma_{ik} = n_i \delta_{ik}. \label{eq:1rdm-nof} 
\end{equation}
Here, $n_i$ represents the occupation number (ON) of the natural spin orbital $\phi_i$. The existence theorems\cite{Gilbert1975, Valone1980} of the NOF do not provide a general analytical expression. While the Levy formulation\cite{Levy1979} can reveal the exact functional for specific model systems, it remains impractical for broader computational applications. This limitation has spurred the development of approximate NOFs.

To date, the only functional known\cite{Husimi1940} with certainty to be exact for a system with at most two-body interactions also involves the two-particle density matrix ($\mathbf{D}$):
\begin{equation}
  E=\sum\limits_{ik} \Gamma_{ik} H_{ki} + \sum\limits_{ijkl} D_{ij,kl} \left\langle kl|ij \right\rangle.
  \label{Eel}
\end{equation}
In this expression, $H_{ki}$ represents the matrix elements of the one-particle part of the Hamiltonian, which includes both kinetic and potential energy operators, while $\left\langle kl|ij \right\rangle$ denotes the two-particle interaction matrix elements. Since the non-interacting part of (\ref{Eel}) explicitly depends on $\mathbf{\Gamma}$, we can obtain the target NOF by expressing the entire component that depends on $\mathbf{D}$ in terms of $\mathbf{\Gamma}$. Unfortunately, analytically determining the electron-electron potential energy in terms of $\mathbf{\Gamma}$ remains a formidable challenge. Therefore, we use the exact functional form (\ref{Eel}) in conjunction with a reconstruction functional $D[n_{i}, n_{j}, n_{k}, n_{l}]$, as follows:
\begin{equation}
  E\left[\left\{ n_{i},\phi_{i}\right\} \right] = \sum\limits _{i}n_{i}H_{ii} + \sum\limits_{ijkl}D[n_{i},n_{j},n_{k},n_{l}]\left\langle kl|ij\right\rangle \label{ENOF}
\end{equation}
in order to obtain an approximate NOF. Restricting the ONs to the interval $[0,1]$ is both a necessary and sufficient condition for ensemble $\mathbb{N}$-representability of $\mathbf{\Gamma}$ within Löwdin normalization, where its trace corresponds to $\mathbb{N}$, the total number of electrons. \cite{Coleman1963} 

A systematic application of an explicit two-index simplification \cite{Piris2006-kn} of the two-particle cumulant \cite{Mazziotti1998, Kutzelnigg1999} has led to the development of the Piris series \cite{Leiva2005, Piris2007a, Piris2010, Piris2010a, Piris2011, Piris2014c, Piris2017} of JKL-only functionals:
\begin{multline}    
    E\left[\left\{ n_{p},\varphi_{p}\right\} \right] = 2 \sum_p n_p H_{pp} + \sum_{pq} A[n_p,n_q] J_{qp}\\ - \sum_{pq} B[n_p,n_q] K_{qp} - \sum_{pq} C[n_p,n_q] L_{qp}
\end{multline}
where $J_{qp}$, $K_{qp}$, and $L_{qp}$ represent the usual Coulomb, exchange, and exchange-time-inversion integrals,\cite{Piris1999} respectively. Note that the indices have been changed to indicate natural spatial orbitals $\left\{\varphi_p\right\}$, and their occupancies $\left\{n_p\right\}$ determine the functions $A$, $B$, and $C$. These NOFs, particularly those based on electron pairing, \cite{Piris2018e} have been successful in capturing static electron correlation. \cite{Mitxelena2019} Dynamic electron correlation can be addressed through perturbative corrections, \cite{Piris2018b, Rodriguez-Mayorga2021} which improve energies but not the quality of natural orbitals (NOs). Achieving fully correlated $\left\{n_p,\varphi_p\right\}$ requires a more comprehensive approach, such as the global NOF (GNOF), \cite{Piris2021b} which balances dynamic and static correlations and shows strong agreement with accurate wavefunction-based methods in both relative and absolute energies.\cite{Mitxelena2024} A review of the current state of the art in NOF approximations can be found elsewhere.\cite{Piris2024a,Piris2024b} Recent advancements in applying machine learning techniques to 1RDM functional theories have demonstrated their potential to refine functional approximations and capture complex correlation effects.\cite{Wetherell2020-nw,Schmidt2021-gp,Shao2023-la} This work aligns with these developments by introducing advanced optimization strategies into NOF theory.

NOFs exhibit a formal fifth-order scaling cost, which can be reduced to the fourth-order.\cite{Lew-Yee2021-bb, Lemke2022-vw} This favorable scalability allows the correlation of a vast number of electrons, surpassing the limitations of small active spaces in traditional multireference approaches. As a result, NOFs enable comprehensive all-electron calculations across the full range of available orbitals. However, the widespread adoption of NOFs has been hindered by the slow optimization process of NOs, restricting their use to systems with a limited number of correlated electrons. The central objective of this work is to overcome this limitation by leveraging novel optimization techniques developed within the deep learning framework to enhance convergence. This strategy has enabled us to present, in this work, the largest NOF calculations to date, thereby paving the way for effectively correlating a large number of electrons.

Minimizing $E\left[\left\{n_{p},\varphi_{p}\right\}\right]$ requires satisfying the orthonormality constraints for the NOs, while the ONs must meet the ensemble $\mathbb{N}$-representability conditions. For electron-pairing-based NOFs, additional conditions are imposed, ensuring the $\mathbf{\Gamma}$ normalization. This approach leads to a challenging constrained optimization problem, typically addressed by optimizing the energy separately for ONs and NOs, as simultaneous optimization has proven to be inefficient.\cite{Cances2008} Furthermore, separate optimization allows fully optimizing the ONs at a feasible cost while focusing computational resources on the more demanding task of NO optimization. Indeed, we have observed that the ONs quickly approximate their final values within the first few external iterations when optimized separately from the NOs, leaving the latter as the real challenge for the optimization, as can be seen in the example provided in the Supplemental Material.

Although advances in optimizing ONs have increased efficiency,\cite{Yao2021-gm,Franco2024-sh} significant improvements in NO optimization are still needed to make NOF-based methods competitive with their density functional-based counterparts. For fixed ONs, optimizing NOs under orthonormality constraints can be performed using the Lagrange multiplier technique as follows:
\begin{equation}
    \Omega\left[\left\{\varphi_{p}\right\}\right] = E\left[\left\{\varphi_{p}\right\}\right] - 2\sum_{pq} \lambda_{qp} (\braket{\varphi_{p}}{\varphi_{q}} - \delta_{pq})
\end{equation}
In the standard self-consistent field approach,\cite{aszabo82:qchem} orbital optimization is typically performed in the canonical orbital representation, where $\mathbf{\lambda}$ is diagonal. However, $\mathbf{\Gamma}$ and $\mathbf{\lambda}$ matrices cannot be simultaneously diagonalized,\cite{Piris2013} except in the special case of Hartree-Fock (HF). A significant consequence of the discrepancy between canonical and NO representations is the impossibility of obtaining a Fockian in NOF calculations. \cite{Piris2013} This issue can be addressed by leveraging the symmetry of $\mathbf{\lambda}$ at minimum and constructing a generalized pseudo-Fockian with well-defined off-diagonal elements, which leads to the iterative diagonalization method.\cite{Piris2009-jo} While this procedure marked a significant advance, its efficiency is constrained by the lack of an exact prescription for the diagonal elements and the absence of effective techniques to improve the convergence rate.

A well-established alternative to diagonalizing is orbital rotation. In this method, orthonormal NOs are rotated using a unitary transformation $\mathbf{U}=e^{\mathbf{y}}$, where $y_{pq}$ is an antisymmetric matrix ($y_{pq} = -y_{qp}$). This transformation makes the energy a function of the $N_{bf}(N_{bf}-1)/2$ independent rotation parameters ($y_{pq}$ with $p<q$), with $N_{bf}$ the number of basis functions, facilitating the use of unrestricted optimization methods.\cite{Helmich-Paris2021-mj, Elayan2022, Cartier2024} Usually, trust-region based methods are used, but these come with the drawback of an increased computational cost due to the Hessian calculation. Gradient-based methods can also be used, but derivatives have to be evaluated at $\mathbf{y}=\mathbf{0}$ forcing one to follow a $\mathbf{y}_{t}=\Delta \mathbf{y}$ update rule, contrasting with the most common step given as $\mathbf{y}_{t} = \mathbf{y}_{t-1} + \Delta \mathbf{y}$. Generally, gradient-based techniques leverage past-step data to approximate the curvature of the surface and refine their subsequent moves. Nevertheless, due to the update rule of the orbital rotation, the optimization landscape shifts after every step, resetting to the origin. This continuous alteration hinders the transfer of gradient data across iterations, thereby reducing the efficiency of standard methods such as conjugate gradients.

To effectively correlate a large number of electrons, a robust convergence scheme with manageable costs is essential. Consequently, we disregard the Hessian and focus on optimizing orbitals through their rotations, using only the orbital gradient. Achieving our goals requires moving beyond traditional methods and adopting innovative optimization techniques. In this context, it is important to underscore the parallelism between the external iteration of our nested iterative method performed in the NOF calculations and the epochs of the neural networks training process. This similarities allow us to drawn inspiration from the deep learning field and apply the adaptive momentum (ADAM) procedure to the optimization of the NOs. This method adjusts each step component based on first and second moments derived solely from the orbital gradients, ensuring efficient scaling and minimal memory usage.\cite{Kingma2014-xg} Additionally, it features step-length annealing, akin to scaling pseudo-Fockian matrix elements in the iterative diagonalization method.

For fixed ONs, the procedure involves calculating the orbital gradient $\mathbf{g}$ at each $t$-th iteration. Since $\mathbf{y}$ is an antisymmetric matrix, only $N_{bf}(N_{bf}-1)/2$ elements are needed, allowing the gradient elements to be computed as:
\begin{equation}
    g_{pq} = \frac{dE}{dy_{pq}}\bigg\rvert_{\mathbf{y}=\mathbf{0}} = 4(\lambda_{pq}-\lambda_{qp}) \>, \>\>  p<q
\end{equation}
where the elements correspond to the vectorization of the 
energy derivatives with respect to the independent orbital rotation parameters, evaluated at $\mathbf{y}=\mathbf{0}$. Next, the element-wise first moment ($\mathbf{m}$) and second moment ($\mathbf{v}$) at the $t$-th iteration are determined as:
\begin{eqnarray}
\mathbf{m}_t &=& \beta_1 \mathbf{m}_{t-1} + (1-\beta_1) \mathbf{g}_t\\
\mathbf{v}_t &=& \beta_2 \mathbf{v}_{t-1} + (1-\beta_2) \mathbf{g}_t^2    
\end{eqnarray}
where $\beta_1$ and $\beta_2$ are the hyperparameters for the moment estimation. These hyperparameters control how quickly the algorithm ``forgets" past gradient information, 
balancing responsiveness to recent gradients with stability from accumulated gradient history. The bias-corrected moments are computed as:
\begin{eqnarray}
    \hat{\mathbf{m}}_t &=& \mathbf{m}_t/(1-\beta_1^t)\\
    \hat{\mathbf{v}}_t &=& \mathbf{v}_t/(1-\beta_2^t)
\end{eqnarray}
In deep learning, taking the maximum of the second moment, i.e., $\hat{\mathbf{v}}_t^{\max}=\max(\hat{\mathbf{v}}_{t-1}^{\max},\hat{\mathbf{v}}_t)$, is optional. However, for the NO optimization, this practice is essential for improving convergence. The step update is then given by:
\begin{equation}
    \mathbf{y}_t=\frac{\alpha \hat{\mathbf{m}_t}}{\sqrt{\hat{\mathbf{v}}^\text{max}_{t} + \epsilon}} \>, \>\>\> \mathbf{U}_t=e^{\mathbf{y}_t} \>, \>\>\> \mathbf{C}_t=\mathbf{C}_{t-1}\mathbf{U}_t
    \label{step}
\end{equation}
where $\alpha$ is the learning rate and $\epsilon = 10^{-16}$. Notably, Eq. ({\ref{step}}) highlights a key difference from the standard ADAM implementation, as $\mathbf{y}_{t-1}$ is set to zero when updating $\mathbf{y}_t$ during optimization, as a consequence  of having the derivatives at $\mathbf{y} = \mathbf{0}$. The new NOs are then used to compute the energy, and the process is repeated until convergence is achieved or early stopping criteria are met. Although avoiding the Hessian leads to a lack of a trust-region, ADAM provides approximately bounded steps through the value of the hyperparameters at only the cost of the gradient calculation.\cite{Kingma2014-xg} Additional details on our ADAM implementation for NO optimization and its integration with ON optimization are provided in the supporting information.

To validate the effectiveness of the proposed method, we present three examples involving a large number of electrons: (i) the symmetric dissociation of a large hydrogen cluster, (ii) an analysis of the ON distribution in fullerenes, and (iii) a study of the singlet-triplet energy gap in linear acenes. Calculations were performed using the DoNOF software\cite{Piris2021-xo} with STO-3G\cite{Hehre1969-es} basis set for the hydrogen cluster and cc-pVDZ \cite{Dunning1989-mo} basis set for the fullerenes and the acenes. Additional analysis on the convergence of the method with respect to the basis set size is provided in the Supplemental Material. Our calculations are performed using real orbitals within a spin-restricted framework. In particular, the functionals PNOF7\cite{Piris2017} and GNOF\cite{Piris2021b} were employed as required by the studied problems.

The study of hydrogen clusters dissociating into individual atoms illustrates a shift from weak to strong correlation, leading to a metal-to-insulator transition. Specifically, we took a 10$\times$10$\times$10 hydrogen cube and simultaneously stretched all internuclear distances between adjacent atoms, resulting in 1000 isolated H atoms. The observed electron localization upon symmetric dissociation makes this system an ideal model for strongly correlated Mott insulators and represents the largest calculation performed within NOF theory to date. This calculation places us in the domain of thousands of correlated electrons, a scenario previously surpassed by a limited number of methods, including Fixed-Node Diffusion Monte Carlo (FN-DMC) in cells of \ce{TiO2},\cite{Luo2016-zm} cells of Para-Diiodobenzene,\cite{Hongo2015-xt} or in the 1AMB protein.\cite{Scemama2013-kd} However, in contrast to these previous calculations, the correlation phenomenon observed in the hydrogen cube involves all the electrons, which represents a unique challenge for electronic structure calculations. 

The symmetric dissociation energy curve for PNOF7 is presented in Fig. 1a. It has been demonstrated that PNOF7 is sufficient to model the potential energy curve for this type of system in smaller cases. \cite{Mitxelena2022-wt} Indeed, PNOF7 successfully captures the expected behavior of the potential energy curve in both the bonding and dissociated regions, resulting in a stable dissociation energy.
\begin{figure}[htb]
    \input{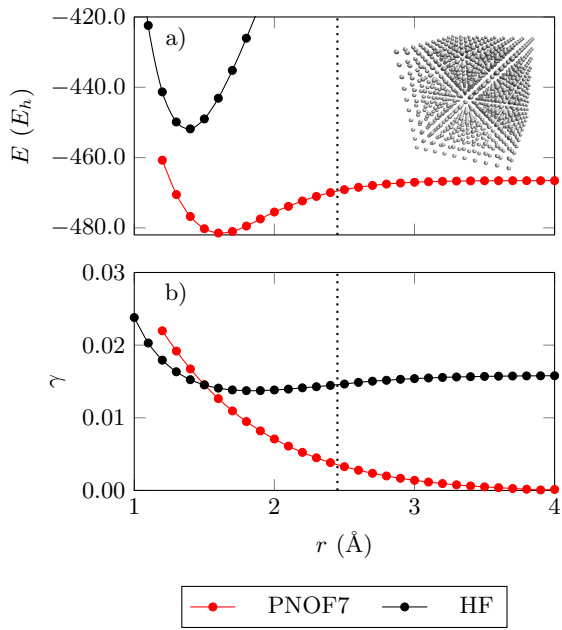}
    \caption{Metal-to-insulator transition of a 1000 Hydrogen atom cluster (10$\times$10$\times$10 Hydrogen cube). a) Potential energy curve of the symmetric dissociation. b) Harmonic average $\gamma$ of the off-diagonal elements of $\Gamma^\text{AO}$. The expected critical distance ($r_c$) for the metal-to-insulator transition is indicated by a vertical dotted line.}
    \label{fig:h_cube}
\end{figure}
To quantitatively describe the metal-to-insulator transition observed in large systems, we computed the harmonic average $\gamma$ of all off-diagonal elements $\Gamma^\text{AO}_{\mu\nu}$ in the atomic orbital basis set. This measure is given by:\cite{Sinitskiy2010-ik}
\begin{equation}
\gamma = \sqrt {\dfrac{1}{N_{bf}\left(N_{bf}-1\right)} \sum\limits _{\mu\neq \nu} (\Gamma ^{AO}_{\mu\nu})^{2} }\label{eq:harmonic-average}
\end{equation}
Fig. \ref{fig:h_cube}b shows the dependence of $\gamma$ on the H–H distances ($r$) for the 10$\times$10$\times$10 H cube, with  restricted HF values provided for comparison. As the interatomic separation increases, the loss of spatial correlation causes the off-diagonal elements $\Gamma^{AO}_{\mu\nu}$ to approach zero, resulting in a harmonic average $\gamma$ that also approaches zero. This behavior demonstrates that PNOF7 effectively captures the metal-to-insulator transition in large 3D systems. 

The interatomic distance at which this transition occurs, $r_c$, corresponds to the critical density $n_c\thicksim 0.01 a_0^{-3}$, where the Coulomb interaction dominates over electron delocalization.\cite{mott1990} Using $n=1/r_c^3$, the critical interatomic distance is found to be $r_c \approx 2.46$ \AA, corresponding to a density of $n_c \approx 0.067 e^-/$\AA$^3$.
At this critical density, the off-diagonal elements of the 1RDM are effectively close to zero ($\left |\Gamma^{AO}_{\mu\nu} \right | < 0.05$), signaling a transition from a metallic state, characterized by delocalized electrons and non-zero off-diagonal elements, to an insulating state, where electrons become localized. This behavior is consistent with the observed trend of $\gamma$ in Fig. \ref{fig:h_cube}b,  where the harmonic average $\gamma$ approaches zero at the transition. Overall, this calculation, involving 1000 electrons in 1000 orbitals, highlights the capability of the proposed approach to accurately capture the metal-to-insulator transition in strongly correlated systems of unprecedented size.

\begin{figure}[htb]
    \input{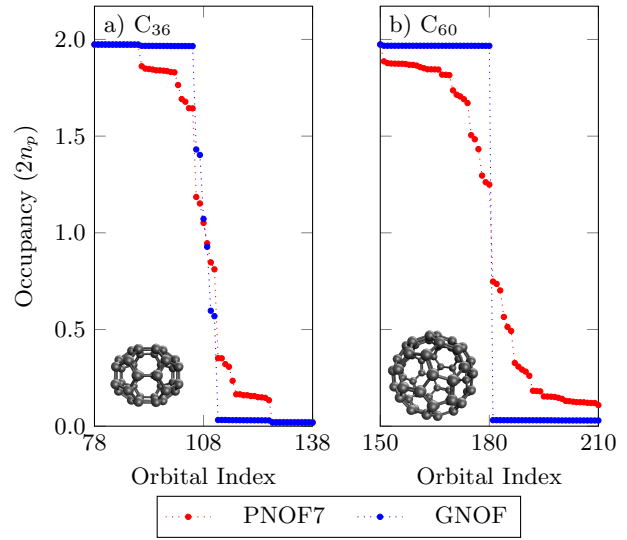}
    \caption{ONs of fullerenes: a) \ce{C36}-D$_\text{6h}$ and b) \ce{C60}. For clarity, only the 30 ONs closest to the Fermi level (below and above) are shown. Geometries were taken from Ref.~\cite{TomanekUnknown-hc}.}
    \label{fig:fullerenes-on}
\end{figure}

The ON distributions of \ce{C36} in D$_\text{6h}$ symmetry and \ce{C60} are presented in Fig.~\ref{fig:fullerenes-on}, highlighting an important feature of NOFs: their ability to provide insight into both static and dynamic electron correlation within a system. Specifically, the presence of fractional ONs indicates static correlation. Although the most common fullerene, \ce{C60}, is typically considered of single-reference character, \ce{C36} is known to exhibit static electron correlation.\cite{Varganov2002-dt,Lee2019-ju,Lee2020-vg} PNOF7 predicts a singlet ground state with fractional occupations for both molecules, as denoted by the red marks. While this result might initially seem unexpected, it can be attributed to the functional's known tendency to overstabilize solutions with fractional occupations due to a lack of dynamic correlation. \cite{Piris2017}

Interestingly, when using GNOF starting from these solutions, fractional occupations are retained for \ce{C36}, as indicated by the blue marks in Fig.~\ref{fig:fullerenes-on}a, whereas they are correctly eliminated for \ce{C60}, as shown in Fig.~\ref{fig:fullerenes-on}b. Notably, with \ce{C60} containing 900 spatial NOs, these molecules exemplify extensive NOF calculations.

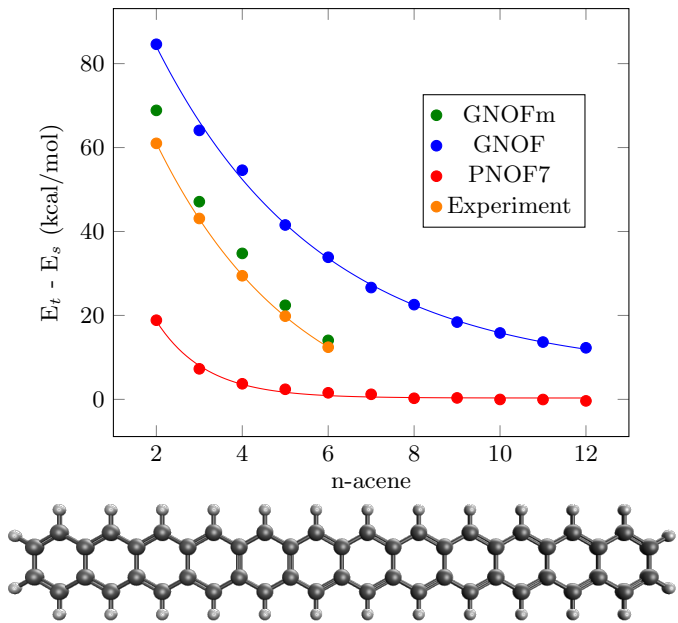
\begin{figure}[htb]
  \centering
  \begin{tikzpicture}
    \begin{axis}[
      xlabel={n-acene},
      ylabel={E$_t$ - E$_s$ (kcal/mol)},
      legend style={at={(0.6,0.8)}, anchor=north west}
    ]
    \addplot[
      %scatter,
      only marks,
      mark=*,
      mark options={Green},
    ]
    coordinates {
    (2, 68.85874954498173)
    (3, 47.10172445999155)
    (4, 34.77785067000585)
    (5, 22.4201734950128)
    (6, 14.059824990023367)
    %(7, 5.072765490040524)
    %(8, 0.47882902511009884)
    %(9, -2.221804305015951)
    %(10, -4.58139346510469)
    %(11, -5.739739514960797)               
    %(12, -5.668056269992053)
    };    
    \addlegendentry{GNOFm}
%    \addplot[
%      domain=2:10,
%      samples=100,
%      color=Green
%    ]{143.64055901*exp(-0.27271118*x) - 14.76269182};
%    \addlegendentry{$143.64\exp(-0.27x) - 17.76$}
    \addplot[
      %scatter,
      only marks,
      mark=*,
      mark options={blue},
    ]
    coordinates {
    (2, 84.61125238500942)
    (3, 64.08927379497402)
    (4, 54.60312832499639)
    (5, 41.54994180004817)
    (6, 33.861020084946375)
    (7, 26.64195915011279)
    (8, 22.57112849990974)
    (9, 18.41124255007413)
    (10, 15.824938664951752)    
    (11, 13.655187810018276)    
    (12, 12.284300625107004)
    };
    \addlegendentry{GNOF}
    \addplot[
      domain=2:12,
      samples=100,
      color=blue,
      forget plot
    ]{131.21005503*exp(-0.2610352*x) + 6.22314722}; 
%    \addlegendentry{$131.21\exp(-0.26x) + 6.22$}    
    \addplot[
      %scatter,
      only marks,
      mark=*,
      mark options={red},
    ]
    coordinates {
      (2, 18.8622888299977)
      (3, 7.2750027150060586)
      (4, 3.712100849967976)
      (5, 2.404555815039379)
      (6, 1.5616035000500574)
      (7, 1.208706195021284)
      (8, 0.2615842650311606)
      (9, 0.3458732248753904)
      (10, -0.0196297950288681)
      (11, -0.034932255047272064)
      (12, -0.3629944200531895)
    };    
    \addlegendentry{PNOF7}
    \addplot[
      domain=2:12,
      samples=100,
      color=red,
      forget plot
    ]{103.24871737*exp(-0.86454423*x) + 0.3236878};
%    \addlegendentry{$103.25\exp(-0.86x) + 0.32$}    
%    \addplot[
%      %scatter,
%      only marks,
%      mark=square*,
%      mark options={orange},
%    ]
%    coordinates {
%    (2, 63.8)
%    (3, 45.2)
%    (4, 32.8)
%    (5, 24.5)
%    (6, 19.7)
%    (8, 15.4)
%    (10, 13.0)
%    (12, 11.9)
%    };
%   \addlegendentry{v2RDM}
    \addplot[
      %scatter,
      only marks,
      mark=*,
      mark options={orange},
    ]
    coordinates {
    (2, 61.00)
    (3, 43.10)
    (4, 29.45)
    (5, 19.83)
    (6, 12.43)
    };
    \addlegendentry{Experiment}
    \addplot[
      domain=2:6,
      samples=100,
      color=orange,
      forget plot
    ]{126.7660853*exp(-0.30178721*x) + -8.28997049};
%    \addlegendentry{$126.77\exp(-0.30x) - 8.29$}   
    \end{axis}
  \end{tikzpicture}
  \includegraphics[width=0.5\textwidth]{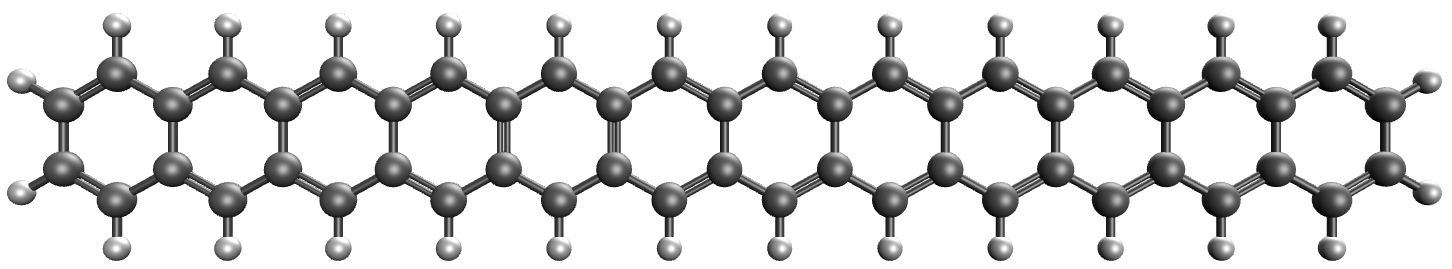}
  \caption{Singlet-triplet energy gap in the linear n-acene series. Geometries were taken from Ref.\cite{Mullinax2019-km}. Experimental data is from Refs.\cite{Marian2008-hz, Grimme2003-cw, Grimme1999-hm, Angliker1982-lq, Burgos1977-ff} and compiled in Ref.~\cite{Hajgato2011-qw}.}
  \label{fig:linear-acenes}
\end{figure}

Regarding the singlet-triplet energy gap of linear acenes, a transition from dynamic to static correlation dominance as the number of rings in the series increases is well established, causing the singlet-triplet energy gap to decay toward a certain limit. \cite{Sony2007-hu, Yang2016-ql, Dupuy2018-nc} However, the theoretical prediction of the curve’s decay and the numerical value of this limit differ depending on the method used. This variation arises from the approximations necessary to make common multireference methods computationally feasible for larger systems. Fig.~\ref{fig:linear-acenes} displays the values calculated with PNOF7 and GNOF, along with experimental values marked in orange.~\cite{Hajgato2011-qw, Marian2008-hz, Grimme2003-cw, Grimme1999-hm, Angliker1982-lq, Burgos1977-ff}

It can be seen that PNOF7 captures the decaying trend in the singlet-triplet energy gap, but the shape of the curve deviates from the experimental results, particularly at the beginning where the systems are predominantly single-reference. These discrepancies decrease as the number of rings increases and static correlation becomes more significant. In contrast, GNOF corrects the predictions at the beginning of the curve by better accounting for dynamic correlation, while still maintaining the overall decaying trend. Fitting the data to a function of the form ($ae^{-bx} + c$) yields an estimated singlet-triplet energy gap for acenes in the limit of infinite rings: 0.32 kcal/mol with PNOF7 and 6.22 kcal/mol with GNOF.

While GNOF produces a qualitatively correct curve shape, it appears to be offset relative to the experimental values. In this work, we have modified the GNOF formulation (see Supplemental Material) to include interactions between strongly occupied orbitals in the anti-parallel spin blocks, resulting in what we call ``modified GNOF" (GNOFm), represented by the green marks. Interestingly, these terms appear to be crucial for predictions that align more closely with the experimental data. However, further investigation is required, which will be addressed in future work.

In summary, NOF theory, combined with modern optimization strategies applied to NOs and ONs, presents a promising approach for studying systems with a large number of correlated electrons. In this study, we achieved satisfactory results using \textit{adaptive momentum}, although other optimization methods used in the deep learning framework could also be explored.\cite{Zhuang2020-of, Zaheer2018-ek} Furthermore, the field remains open to the development of new optimization strategies for NOFs, particularly by leveraging advancements in deep learning optimization techniques. The methodology presented in this work, although developed for fermionic systems, could potentially be extended to bosonic systems or mixtures of bosons and fermions.\cite{Schmidt2019-gs} We anticipate that the current implementation will significantly broaden the applicability of NOF theory, allowing for the full utilization of its efficient arithmetic scaling and physical insights. Additionally, this will enhance the accuracy of new functionals, as more comprehensive testing can be conducted throughout their development.

\begin{acknowledgments}
\textit{Acknowledgments} - J. F. H. Lew-Yee acknowledges the Donostia International Physics Center (DIPC) and the MCIN program ``Severo Ochoa'' under reference AEI/ CEX2018-000867-S for post-doctoral funding (Ref.: 2023/74.) J. M. del Campo acknowledges funding with project Grant No. IN201822 from PAPIIT, and computing resources from ``Laboratorio Nacional de Cómputo de Alto Desempeño (LANCAD)'' with project Grant No. LANCAD-UNAMDGTIC-270. M. Piris acknowledges funding from MCIN/AEI/10.13039/501100011033 (Ref.: PID2021-126714NB-I00) and the Eusko Jaurlaritza (Ref.: IT1584-22).
\end{acknowledgments}

%\bibliography{references}

\begin{thebibliography}{66}%
\makeatletter
\providecommand \@ifxundefined [1]{%
 \@ifx{#1\undefined}
}%
\providecommand \@ifnum [1]{%
 \ifnum #1\expandafter \@firstoftwo
 \else \expandafter \@secondoftwo
 \fi
}%
\providecommand \@ifx [1]{%
 \ifx #1\expandafter \@firstoftwo
 \else \expandafter \@secondoftwo
 \fi
}%
\providecommand \natexlab [1]{#1}%
\providecommand \enquote  [1]{``#1''}%
\providecommand \bibnamefont  [1]{#1}%
\providecommand \bibfnamefont [1]{#1}%
\providecommand \citenamefont [1]{#1}%
\providecommand \@href[1]{\@@startlink{#1}\@@href}%
\providecommand \@@href[1]{\endgroup#1\@@endlink}%
\providecommand \@sanitize@url [0]{\catcode `\\12\catcode `\$12\catcode `\&12\catcode `\#12\catcode `\^12\catcode `\_12\catcode `\%12\relax}%
\providecommand \@@startlink[1]{}%
\providecommand \@@endlink[0]{}%
\providecommand \@url [1]{\endgroup\@href {#1}{\urlprefix }}%
\providecommand \urlprefix  [0]{URL }%
\providecommand \doibase [0]{https://doi.org/}%
\providecommand \selectlanguage [0]{\@gobble}%
\providecommand \bibinfo  [0]{\@secondoftwo}%
\providecommand \bibfield  [0]{\@secondoftwo}%
\providecommand \translation [1]{[#1]}%
\providecommand \BibitemOpen [0]{}%
\providecommand \bibitemStop [0]{}%
\providecommand \bibitemNoStop [0]{.\EOS\space}%
\providecommand \EOS [0]{\spacefactor3000\relax}%
\providecommand \BibitemShut  [1]{\csname bibitem#1\endcsname}%
\let\auto@bib@innerbib\@empty
%</preamble>
\bibitem [{\citenamefont {L{\"{o}}wdin}(1955)}]{Lowdin1955a}%
  \BibitemOpen
  \bibfield  {author} {\bibinfo {author} {\bibfnamefont {P.~O.}\ \bibnamefont {L{\"{o}}wdin}},\ }\href {https://doi.org/10.1103/PhysRev.97.1474} {\bibfield  {journal} {\bibinfo  {journal} {Phys. Rev.}\ }\textbf {\bibinfo {volume} {97}},\ \bibinfo {pages} {1474} (\bibinfo {year} {1955})}\BibitemShut {NoStop}%
\bibitem [{\citenamefont {Gilbert}(1975)}]{Gilbert1975}%
  \BibitemOpen
  \bibfield  {author} {\bibinfo {author} {\bibfnamefont {T.~L.}\ \bibnamefont {Gilbert}},\ } {\bibfield  {journal} {\bibinfo  {journal} {Phys. Rev.}\ }\textbf {\bibinfo {volume} {12}},\ \bibinfo {pages} {2111} (\bibinfo {year} {1975})}\BibitemShut {NoStop}%
\bibitem [{\citenamefont {Valone}(1980)}]{Valone1980}%
  \BibitemOpen
  \bibfield  {author} {\bibinfo {author} {\bibfnamefont {S.~M.}\ \bibnamefont {Valone}},\ }\href {https://doi.org/10.1063/1.440249} {\bibfield  {journal} {\bibinfo  {journal} {J. Chem. Phys.}\ }\textbf {\bibinfo {volume} {73}},\ \bibinfo {pages} {1344} (\bibinfo {year} {1980})}\BibitemShut {NoStop}%
\bibitem [{\citenamefont {Levy}(1979)}]{Levy1979}%
  \BibitemOpen
  \bibfield  {author} {\bibinfo {author} {\bibfnamefont {M.}~\bibnamefont {Levy}},\ }\href {https://doi.org/10.1088/0022-3719/12/3/015} {\bibfield  {journal} {\bibinfo  {journal} {Proc. Natl. Acad. Sci. USA}\ }\textbf {\bibinfo {volume} {76}},\ \bibinfo {pages} {6062} (\bibinfo {year} {1979})}\BibitemShut {NoStop}%
\bibitem [{\citenamefont {Husimi}(1940)}]{Husimi1940}%
  \BibitemOpen
  \bibfield  {author} {\bibinfo {author} {\bibfnamefont {K.}~\bibnamefont {Husimi}},\ }\href {https://doi.org/10.11429/ppmsj1919.22.4_264} {\bibfield  {journal} {\bibinfo  {journal} {Proc. Phys. Math. Soc. Jpn.}\ }\textbf {\bibinfo {volume} {22}},\ \bibinfo {pages} {264} (\bibinfo {year} {1940})}\BibitemShut {NoStop}%
\bibitem [{\citenamefont {Coleman}(1963)}]{Coleman1963}%
  \BibitemOpen
  \bibfield  {author} {\bibinfo {author} {\bibfnamefont {A.~J.}\ \bibnamefont {Coleman}},\ }\href {https://doi.org/10.1103/RevModPhys.35.668} {\bibfield  {journal} {\bibinfo  {journal} {Rev. Mod. Phys.}\ }\textbf {\bibinfo {volume} {35}},\ \bibinfo {pages} {668} (\bibinfo {year} {1963})}\BibitemShut {NoStop}%
\bibitem [{\citenamefont {Piris}(2006)}]{Piris2006-kn}%
  \BibitemOpen
  \bibfield  {author} {\bibinfo {author} {\bibfnamefont {M.}~\bibnamefont {Piris}},\ } {\bibfield  {journal} {\bibinfo  {journal} {Int. J. Quantum Chem.}\ }\textbf {\bibinfo {volume} {106}},\ \bibinfo {pages} {1093} (\bibinfo {year} {2006})}\BibitemShut {NoStop}%
\bibitem [{\citenamefont {Mazziotti}(1998)}]{Mazziotti1998}%
  \BibitemOpen
  \bibfield  {author} {\bibinfo {author} {\bibfnamefont {D.~A.}\ \bibnamefont {Mazziotti}},\ }\href {https://doi.org/10.1016/S0009-2614(98)00470-9} {\bibfield  {journal} {\bibinfo  {journal} {Chem. Phys. Lett.}\ }\textbf {\bibinfo {volume} {289}},\ \bibinfo {pages} {419} (\bibinfo {year} {1998})}\BibitemShut {NoStop}%
\bibitem [{\citenamefont {Kutzelnigg}\ and\ \citenamefont {Mukherjee}(1999)}]{Kutzelnigg1999}%
  \BibitemOpen
  \bibfield  {author} {\bibinfo {author} {\bibfnamefont {W.}~\bibnamefont {Kutzelnigg}}\ and\ \bibinfo {author} {\bibfnamefont {D.}~\bibnamefont {Mukherjee}},\ }\href {https://doi.org/10.1063/1.478189} {\bibfield  {journal} {\bibinfo  {journal} {J. Chem. Phys.}\ }\textbf {\bibinfo {volume} {110}},\ \bibinfo {pages} {2800} (\bibinfo {year} {1999})}\BibitemShut {NoStop}%
\bibitem [{\citenamefont {Leiva}\ and\ \citenamefont {Piris}(2005)}]{Leiva2005}%
  \BibitemOpen
  \bibfield  {author} {\bibinfo {author} {\bibfnamefont {P.}~\bibnamefont {Leiva}}\ and\ \bibinfo {author} {\bibfnamefont {M.}~\bibnamefont {Piris}},\ }\href {https://doi.org/10.1063/1.2135289} {\bibfield  {journal} {\bibinfo  {journal} {J. Chem. Phys.}\ }\textbf {\bibinfo {volume} {123}},\ \bibinfo {pages} {214102} (\bibinfo {year} {2005})}\BibitemShut {NoStop}%
\bibitem [{\citenamefont {Piris}\ \emph {et~al.}(2007)\citenamefont {Piris}, \citenamefont {Lopez},\ and\ \citenamefont {Ugalde}}]{Piris2007a}%
  \BibitemOpen
  \bibfield  {author} {\bibinfo {author} {\bibfnamefont {M.}~\bibnamefont {Piris}}, \bibinfo {author} {\bibfnamefont {X.}~\bibnamefont {Lopez}},\ and\ \bibinfo {author} {\bibfnamefont {J.~M.}\ \bibnamefont {Ugalde}},\ }\href {https://doi.org/10.1063/1.2743019} {\bibfield  {journal} {\bibinfo  {journal} {J. Chem. Phys.}\ }\textbf {\bibinfo {volume} {126}},\ \bibinfo {pages} {214103} (\bibinfo {year} {2007})}\BibitemShut {NoStop}%
\bibitem [{\citenamefont {Piris}\ \emph {et~al.}(2010{\natexlab{a}})\citenamefont {Piris}, \citenamefont {Matxain}, \citenamefont {Lopez},\ and\ \citenamefont {Ugalde}}]{Piris2010}%
  \BibitemOpen
  \bibfield  {author} {\bibinfo {author} {\bibfnamefont {M.}~\bibnamefont {Piris}}, \bibinfo {author} {\bibfnamefont {J.~M.}\ \bibnamefont {Matxain}}, \bibinfo {author} {\bibfnamefont {X.}~\bibnamefont {Lopez}},\ and\ \bibinfo {author} {\bibfnamefont {J.~M.}\ \bibnamefont {Ugalde}},\ }\href {https://doi.org/10.1063/1.3298694} {\bibfield  {journal} {\bibinfo  {journal} {J. Chem. Phys.}\ }\textbf {\bibinfo {volume} {132}},\ \bibinfo {pages} {031103} (\bibinfo {year} {2010}{\natexlab{a}})}\BibitemShut {NoStop}%
\bibitem [{\citenamefont {Piris}\ \emph {et~al.}(2010{\natexlab{b}})\citenamefont {Piris}, \citenamefont {Matxain}, \citenamefont {Lopez},\ and\ \citenamefont {Ugalde}}]{Piris2010a}%
  \BibitemOpen
  \bibfield  {author} {\bibinfo {author} {\bibfnamefont {M.}~\bibnamefont {Piris}}, \bibinfo {author} {\bibfnamefont {J.~M.}\ \bibnamefont {Matxain}}, \bibinfo {author} {\bibfnamefont {X.}~\bibnamefont {Lopez}},\ and\ \bibinfo {author} {\bibfnamefont {J.~M.}\ \bibnamefont {Ugalde}},\ }\href {https://doi.org/10.1063/1.3481578} {\bibfield  {journal} {\bibinfo  {journal} {J. Chem. Phys.}\ }\textbf {\bibinfo {volume} {133}},\ \bibinfo {pages} {111101} (\bibinfo {year} {2010}{\natexlab{b}})}\BibitemShut {NoStop}%
\bibitem [{\citenamefont {Piris}\ \emph {et~al.}(2011)\citenamefont {Piris}, \citenamefont {Lopez}, \citenamefont {Ruip{\'{e}}rez}, \citenamefont {Matxain},\ and\ \citenamefont {Ugalde}}]{Piris2011}%
  \BibitemOpen
  \bibfield  {author} {\bibinfo {author} {\bibfnamefont {M.}~\bibnamefont {Piris}}, \bibinfo {author} {\bibfnamefont {X.}~\bibnamefont {Lopez}}, \bibinfo {author} {\bibfnamefont {F.}~\bibnamefont {Ruip{\'{e}}rez}}, \bibinfo {author} {\bibfnamefont {J.~M.}\ \bibnamefont {Matxain}},\ and\ \bibinfo {author} {\bibfnamefont {J.~M.}\ \bibnamefont {Ugalde}},\ }\href {https://doi.org/10.1063/1.3582792} {\bibfield  {journal} {\bibinfo  {journal} {J. Chem. Phys.}\ }\textbf {\bibinfo {volume} {134}},\ \bibinfo {pages} {164102} (\bibinfo {year} {2011})}\BibitemShut {NoStop}%
\bibitem [{\citenamefont {Piris}(2014)}]{Piris2014c}%
  \BibitemOpen
  \bibfield  {author} {\bibinfo {author} {\bibfnamefont {M.}~\bibnamefont {Piris}},\ }\href {https://doi.org/10.1063/1.4890653} {\bibfield  {journal} {\bibinfo  {journal} {J. Chem. Phys.}\ }\textbf {\bibinfo {volume} {141}},\ \bibinfo {pages} {044107} (\bibinfo {year} {2014})}\BibitemShut {NoStop}%
\bibitem [{\citenamefont {Piris}(2017)}]{Piris2017}%
  \BibitemOpen
  \bibfield  {author} {\bibinfo {author} {\bibfnamefont {M.}~\bibnamefont {Piris}},\ }\href {https://doi.org/10.1103/PhysRevLett.119.063002} {\bibfield  {journal} {\bibinfo  {journal} {Phys. Rev. Lett.}\ }\textbf {\bibinfo {volume} {119}},\ \bibinfo {pages} {063002} (\bibinfo {year} {2017})}\BibitemShut {NoStop}%
\bibitem [{\citenamefont {Piris}(1999)}]{Piris1999}%
  \BibitemOpen
  \bibfield  {author} {\bibinfo {author} {\bibfnamefont {M.}~\bibnamefont {Piris}},\ }\href {https://doi.org/10.1023/A:1019111828412} {\bibfield  {journal} {\bibinfo  {journal} {J. Math. Chem.}\ }\textbf {\bibinfo {volume} {25}},\ \bibinfo {pages} {47} (\bibinfo {year} {1999})}\BibitemShut {NoStop}%
\bibitem [{\citenamefont {Piris}(2018{\natexlab{a}})}]{Piris2018e}%
  \BibitemOpen
  \bibfield  {author} {\bibinfo {author} {\bibfnamefont {M.}~\bibnamefont {Piris}},\ }in\ \href {https://doi.org/10.1201/9781351170963} {\emph {\bibinfo {booktitle} {Quantum Chemistry at the Dawn of the 21st Century. Series: Innovations in Computational Chemistry}}},\ \bibinfo {editor} {edited by\ \bibinfo {editor} {\bibfnamefont {R.}~\bibnamefont {Carb{\'{o}}-Dorca}}\ and\ \bibinfo {editor} {\bibfnamefont {T.}~\bibnamefont {Chakraborty}}}\ (\bibinfo  {publisher} {Apple Academic Press},\ \bibinfo {year} {2018})\ Chap.~\bibinfo {chapter} {22}, pp.\ \bibinfo {pages} {593--620}\BibitemShut {NoStop}%
\bibitem [{\citenamefont {Mitxelena}\ \emph {et~al.}(2019)\citenamefont {Mitxelena}, \citenamefont {Piris},\ and\ \citenamefont {Ugalde}}]{Mitxelena2019}%
  \BibitemOpen
  \bibfield  {author} {\bibinfo {author} {\bibfnamefont {I.}~\bibnamefont {Mitxelena}}, \bibinfo {author} {\bibfnamefont {M.}~\bibnamefont {Piris}},\ and\ \bibinfo {author} {\bibfnamefont {J.~M.}\ \bibnamefont {Ugalde}},\ }in\ \href {https://doi.org/10.1016/bs.aiq.2019.04.001} {\emph {\bibinfo {booktitle} {State Art Mol. Electron. Struct. Comput. Correl. Methods, Basis Sets More}}},\ \bibinfo {series} {Adv. Quantum Chem.}, Vol.~\bibinfo {volume} {79},\ \bibinfo {editor} {edited by\ \bibinfo {editor} {\bibfnamefont {P.}~\bibnamefont {Hoggan}}\ and\ \bibinfo {editor} {\bibfnamefont {U.}~\bibnamefont {Ancarani}}}\ (\bibinfo  {publisher} {Academic Press},\ \bibinfo {year} {2019})\ pp.\ \bibinfo {pages} {155--177}\BibitemShut {NoStop}%
\bibitem [{\citenamefont {Piris}(2018{\natexlab{b}})}]{Piris2018b}%
  \BibitemOpen
  \bibfield  {author} {\bibinfo {author} {\bibfnamefont {M.}~\bibnamefont {Piris}},\ }\href {https://doi.org/10.1103/PhysRevA.98.022504} {\bibfield  {journal} {\bibinfo  {journal} {Phys. Rev. A}\ }\textbf {\bibinfo {volume} {98}},\ \bibinfo {pages} {022504} (\bibinfo {year} {2018}{\natexlab{b}})}\BibitemShut {NoStop}%
\bibitem [{\citenamefont {Rodr{\'{i}}guez-Mayorga}\ \emph {et~al.}(2021)\citenamefont {Rodr{\'{i}}guez-Mayorga}, \citenamefont {Mitxelena}, \citenamefont {Bruneval},\ and\ \citenamefont {Piris}}]{Rodriguez-Mayorga2021}%
  \BibitemOpen
  \bibfield  {author} {\bibinfo {author} {\bibfnamefont {M.}~\bibnamefont {Rodr{\'{i}}guez-Mayorga}}, \bibinfo {author} {\bibfnamefont {I.}~\bibnamefont {Mitxelena}}, \bibinfo {author} {\bibfnamefont {F.}~\bibnamefont {Bruneval}},\ and\ \bibinfo {author} {\bibfnamefont {M.}~\bibnamefont {Piris}},\ }\href {https://doi.org/10.1021/acs.jctc.1c00858} {\bibfield  {journal} {\bibinfo  {journal} {J. Chem. Theory Comput.}\ }\textbf {\bibinfo {volume} {17}},\ \bibinfo {pages} {7562} (\bibinfo {year} {2021})}\BibitemShut {NoStop}%
\bibitem [{\citenamefont {Piris}(2021)}]{Piris2021b}%
  \BibitemOpen
  \bibfield  {author} {\bibinfo {author} {\bibfnamefont {M.}~\bibnamefont {Piris}},\ }\href {https://doi.org/10.1103/PhysRevLett.127.233001} {\bibfield  {journal} {\bibinfo  {journal} {Phys. Rev. Lett.}\ }\textbf {\bibinfo {volume} {127}},\ \bibinfo {pages} {233001} (\bibinfo {year} {2021})}\BibitemShut {NoStop}%
\bibitem [{\citenamefont {Mitxelena}\ and\ \citenamefont {Piris}(2024)}]{Mitxelena2024}%
  \BibitemOpen
  \bibfield  {author} {\bibinfo {author} {\bibfnamefont {I.}~\bibnamefont {Mitxelena}}\ and\ \bibinfo {author} {\bibfnamefont {M.}~\bibnamefont {Piris}},\ }\href {https://doi.org/10.1063/5.0207325} {\bibfield  {journal} {\bibinfo  {journal} {J. Chem. Phys.}\ }\textbf {\bibinfo {volume} {160}},\ \bibinfo {pages} {204106} (\bibinfo {year} {2024})}\BibitemShut {NoStop}%
\bibitem [{\citenamefont {Piris}(2024{\natexlab{a}})}]{Piris2024a}%
  \BibitemOpen
  \bibfield  {author} {\bibinfo {author} {\bibfnamefont {M.}~\bibnamefont {Piris}},\ }in\ \href {https://doi.org/https://doi.org/10.1016/bs.aiq.2024.04.002} {\emph {\bibinfo {booktitle} {Novel Treatments of Strong Correlations}}},\ \bibinfo {series} {Advances in Quantum Chemistry}, Vol.~\bibinfo {volume} {90},\ \bibinfo {editor} {edited by\ \bibinfo {editor} {\bibfnamefont {R.~A.~M.}\ \bibnamefont {Quintana}}\ and\ \bibinfo {editor} {\bibfnamefont {J.~F.}\ \bibnamefont {Stanton}}}\ (\bibinfo  {publisher} {Academic Press},\ \bibinfo {year} {2024})\ pp.\ \bibinfo {pages} {15--66}\BibitemShut {NoStop}%
\bibitem [{\citenamefont {Piris}(2024{\natexlab{b}})}]{Piris2024b}%
  \BibitemOpen
  \bibfield  {author} {\bibinfo {author} {\bibfnamefont {M.}~\bibnamefont {Piris}},\ }\href {https://doi.org/10.1039/D4SC05810K} {\bibfield  {journal} {\bibinfo  {journal} {Chem. Sci.}\ }\textbf {\bibinfo {volume} {15}},\ \bibinfo {pages} {17284} (\bibinfo {year} {2024}{\natexlab{b}})}\BibitemShut {NoStop}%
\bibitem [{\citenamefont {Wetherell}\ \emph {et~al.}(2020)\citenamefont {Wetherell}, \citenamefont {Costamagna}, \citenamefont {Gatti},\ and\ \citenamefont {Reining}}]{Wetherell2020-nw}%
  \BibitemOpen
  \bibfield  {author} {\bibinfo {author} {\bibfnamefont {J.}~\bibnamefont {Wetherell}}, \bibinfo {author} {\bibfnamefont {A.}~\bibnamefont {Costamagna}}, \bibinfo {author} {\bibfnamefont {M.}~\bibnamefont {Gatti}},\ and\ \bibinfo {author} {\bibfnamefont {L.}~\bibnamefont {Reining}},\ } {\bibfield  {journal} {\bibinfo  {journal} {Faraday Discuss.}\ }\textbf {\bibinfo {volume} {224}},\ \bibinfo {pages} {265} (\bibinfo {year} {2020})}\BibitemShut {NoStop}%
\bibitem [{\citenamefont {Schmidt}\ \emph {et~al.}(2021)\citenamefont {Schmidt}, \citenamefont {Fadel},\ and\ \citenamefont {Benavides-Riveros}}]{Schmidt2021-gp}%
  \BibitemOpen
  \bibfield  {author} {\bibinfo {author} {\bibfnamefont {J.}~\bibnamefont {Schmidt}}, \bibinfo {author} {\bibfnamefont {M.}~\bibnamefont {Fadel}},\ and\ \bibinfo {author} {\bibfnamefont {C.~L.}\ \bibnamefont {Benavides-Riveros}},\ } {\bibfield  {journal} {\bibinfo  {journal} {Phys. Rev. Res.}\ }\textbf {\bibinfo {volume} {3}},\ \bibinfo {pages} {L032063} (\bibinfo {year} {2021})}\BibitemShut {NoStop}%
\bibitem [{\citenamefont {Shao}\ \emph {et~al.}(2023)\citenamefont {Shao}, \citenamefont {Paetow}, \citenamefont {Tuckerman},\ and\ \citenamefont {Pavanello}}]{Shao2023-la}%
  \BibitemOpen
  \bibfield  {author} {\bibinfo {author} {\bibfnamefont {X.}~\bibnamefont {Shao}}, \bibinfo {author} {\bibfnamefont {L.}~\bibnamefont {Paetow}}, \bibinfo {author} {\bibfnamefont {M.~E.}\ \bibnamefont {Tuckerman}},\ and\ \bibinfo {author} {\bibfnamefont {M.}~\bibnamefont {Pavanello}},\ } {\bibfield  {journal} {\bibinfo  {journal} {Nat. Commun.}\ }\textbf {\bibinfo {volume} {14}},\ \bibinfo {pages} {6281} (\bibinfo {year} {2023})}\BibitemShut {NoStop}%
\bibitem [{\citenamefont {Lew-Yee}\ \emph {et~al.}(2021)\citenamefont {Lew-Yee}, \citenamefont {Piris},\ and\ \citenamefont {M.~del Campo}}]{Lew-Yee2021-bb}%
  \BibitemOpen
  \bibfield  {author} {\bibinfo {author} {\bibfnamefont {J.~F.~H.}\ \bibnamefont {Lew-Yee}}, \bibinfo {author} {\bibfnamefont {M.}~\bibnamefont {Piris}},\ and\ \bibinfo {author} {\bibfnamefont {J.}~\bibnamefont {M.~del Campo}},\ } {\bibfield  {journal} {\bibinfo  {journal} {J. Chem. Phys.}\ }\textbf {\bibinfo {volume} {154}},\ \bibinfo {pages} {064102} (\bibinfo {year} {2021})}\BibitemShut {NoStop}%
\bibitem [{\citenamefont {Lemke}\ \emph {et~al.}(2022)\citenamefont {Lemke}, \citenamefont {Kussmann},\ and\ \citenamefont {Ochsenfeld}}]{Lemke2022-vw}%
  \BibitemOpen
  \bibfield  {author} {\bibinfo {author} {\bibfnamefont {Y.}~\bibnamefont {Lemke}}, \bibinfo {author} {\bibfnamefont {J.}~\bibnamefont {Kussmann}},\ and\ \bibinfo {author} {\bibfnamefont {C.}~\bibnamefont {Ochsenfeld}},\ } {\bibfield  {journal} {\bibinfo  {journal} {J. Chem. Theory Comput.}\ }\textbf {\bibinfo {volume} {18}},\ \bibinfo {pages} {4229} (\bibinfo {year} {2022})}\BibitemShut {NoStop}%
\bibitem [{\citenamefont {Canc{\`{e}}s}\ and\ \citenamefont {Pernal}(2008)}]{Cances2008}%
  \BibitemOpen
  \bibfield  {author} {\bibinfo {author} {\bibfnamefont {E.}~\bibnamefont {Canc{\`{e}}s}}\ and\ \bibinfo {author} {\bibfnamefont {K.}~\bibnamefont {Pernal}},\ }\href {https://doi.org/10.1063/1.2888550} {\bibfield  {journal} {\bibinfo  {journal} {J. Chem. Phys.}\ }\textbf {\bibinfo {volume} {128}},\ \bibinfo {pages} {134108} (\bibinfo {year} {2008})}\BibitemShut {NoStop}%
\bibitem [{\citenamefont {Yao}\ \emph {et~al.}(2021)\citenamefont {Yao}, \citenamefont {Fang},\ and\ \citenamefont {Su}}]{Yao2021-gm}%
  \BibitemOpen
  \bibfield  {author} {\bibinfo {author} {\bibfnamefont {Y.-F.}\ \bibnamefont {Yao}}, \bibinfo {author} {\bibfnamefont {W.-H.}\ \bibnamefont {Fang}},\ and\ \bibinfo {author} {\bibfnamefont {N.~Q.}\ \bibnamefont {Su}},\ } {\bibfield  {journal} {\bibinfo  {journal} {J. Phys. Chem. Lett.}\ }\textbf {\bibinfo {volume} {12}},\ \bibinfo {pages} {6788} (\bibinfo {year} {2021})}\BibitemShut {NoStop}%
\bibitem [{\citenamefont {Franco}\ \emph {et~al.}(2024)\citenamefont {Franco}, \citenamefont {Bonfil-Rivera}, \citenamefont {Huan Lew-Yee}, \citenamefont {Piris}, \citenamefont {M.~del~Campo},\ and\ \citenamefont {Vargas-Hernández}}]{Franco2024-sh}%
  \BibitemOpen
  \bibfield  {author} {\bibinfo {author} {\bibfnamefont {L.}~\bibnamefont {Franco}}, \bibinfo {author} {\bibfnamefont {I.~A.}\ \bibnamefont {Bonfil-Rivera}}, \bibinfo {author} {\bibfnamefont {J.~F.~H.}\ \bibnamefont {Lew-Yee}}, \bibinfo {author} {\bibfnamefont {M.}~\bibnamefont {Piris}}, \bibinfo {author} {\bibfnamefont {J.}~\bibnamefont {M.~del~Campo}},\ and\ \bibinfo {author} {\bibfnamefont {R.~A.}\ \bibnamefont {Vargas-Hernández}},\ } {\bibfield  {journal} {\bibinfo  {journal} {J. Chem. Phys.}\ }\textbf {\bibinfo {volume} {160}},\ \bibinfo {pages} {244107} (\bibinfo {year} {2024})}\BibitemShut {NoStop}%
\bibitem [{\citenamefont {Szabo}\ and\ \citenamefont {Ostlund}(1996)}]{aszabo82:qchem}%
  \BibitemOpen
  \bibfield  {author} {\bibinfo {author} {\bibfnamefont {A.}~\bibnamefont {Szabo}}\ and\ \bibinfo {author} {\bibfnamefont {N.~S.}\ \bibnamefont {Ostlund}},\ } {\emph {\bibinfo {title} {Modern Quantum Chemistry: Introduction to Advanced Electronic Structure Theory}}},\ \bibinfo {edition} {1st}\ ed.\ (\bibinfo  {publisher} {Dover Publications, Inc.},\ \bibinfo {year} {1996})\BibitemShut {NoStop}%
\bibitem [{\citenamefont {Piris}\ \emph {et~al.}(2013)\citenamefont {Piris}, \citenamefont {Matxain}, \citenamefont {Lopez},\ and\ \citenamefont {Ugalde}}]{Piris2013}%
  \BibitemOpen
  \bibfield  {author} {\bibinfo {author} {\bibfnamefont {M.}~\bibnamefont {Piris}}, \bibinfo {author} {\bibfnamefont {J.~M.}\ \bibnamefont {Matxain}}, \bibinfo {author} {\bibfnamefont {X.}~\bibnamefont {Lopez}},\ and\ \bibinfo {author} {\bibfnamefont {J.~M.}\ \bibnamefont {Ugalde}},\ }\href {https://doi.org/10.1007/s00214-012-1298-4} {\bibfield  {journal} {\bibinfo  {journal} {Theor. Chem. Acc.}\ }\textbf {\bibinfo {volume} {132}},\ \bibinfo {pages} {1298} (\bibinfo {year} {2013})}\BibitemShut {NoStop}%
\bibitem [{\citenamefont {Piris}\ and\ \citenamefont {Ugalde}(2009)}]{Piris2009-jo}%
  \BibitemOpen
  \bibfield  {author} {\bibinfo {author} {\bibfnamefont {M.}~\bibnamefont {Piris}}\ and\ \bibinfo {author} {\bibfnamefont {J.~M.}\ \bibnamefont {Ugalde}},\ } {\bibfield  {journal} {\bibinfo  {journal} {J. Comput. Chem.}\ }\textbf {\bibinfo {volume} {30}},\ \bibinfo {pages} {2078} (\bibinfo {year} {2009})}\BibitemShut {NoStop}%
\bibitem [{\citenamefont {Helmich-Paris}(2021)}]{Helmich-Paris2021-mj}%
  \BibitemOpen
  \bibfield  {author} {\bibinfo {author} {\bibfnamefont {B.}~\bibnamefont {Helmich-Paris}},\ } {\bibfield  {journal} {\bibinfo  {journal} {J. Chem. Phys.}\ }\textbf {\bibinfo {volume} {154}},\ \bibinfo {pages} {164104} (\bibinfo {year} {2021})}\BibitemShut {NoStop}%
\bibitem [{\citenamefont {Elayan}\ \emph {et~al.}(2022)\citenamefont {Elayan}, \citenamefont {Gupta},\ and\ \citenamefont {Hollett}}]{Elayan2022}%
  \BibitemOpen
  \bibfield  {author} {\bibinfo {author} {\bibfnamefont {I.~A.}\ \bibnamefont {Elayan}}, \bibinfo {author} {\bibfnamefont {R.}~\bibnamefont {Gupta}},\ and\ \bibinfo {author} {\bibfnamefont {J.~W.}\ \bibnamefont {Hollett}},\ }\href {https://doi.org/10.1063/5.0073227} {\bibfield  {journal} {\bibinfo  {journal} {J. Chem. Phys.}\ }\textbf {\bibinfo {volume} {156}},\ \bibinfo {pages} {094102} (\bibinfo {year} {2022})}\BibitemShut {NoStop}%
\bibitem [{\citenamefont {Cartier}\ and\ \citenamefont {Giesbertz}(2024)}]{Cartier2024}%
  \BibitemOpen
  \bibfield  {author} {\bibinfo {author} {\bibfnamefont {N.~G.}\ \bibnamefont {Cartier}}\ and\ \bibinfo {author} {\bibfnamefont {K.~J.}\ \bibnamefont {Giesbertz}},\ }\href {https://doi.org/10.1021/acs.jctc.4c00118} {\bibfield  {journal} {\bibinfo  {journal} {J. Chem. Theory Comput.}\ }\textbf {\bibinfo {volume} {20}},\ \bibinfo {pages} {3669} (\bibinfo {year} {2024})}\BibitemShut {NoStop}%
\bibitem [{\citenamefont {Kingma}\ and\ \citenamefont {Ba}(2014)}]{Kingma2014-xg}%
  \BibitemOpen
  \bibfield  {author} {\bibinfo {author} {\bibfnamefont {D.~P.}\ \bibnamefont {Kingma}}\ and\ \bibinfo {author} {\bibfnamefont {J.}~\bibnamefont {Ba}},\ } {\bibfield  {journal} {\bibinfo  {journal} {arXiv:1412.6980 [cs.LG]}\ } (\bibinfo {year} {2014})}\BibitemShut {NoStop}%
\bibitem [{\citenamefont {Piris}\ and\ \citenamefont {Mitxelena}(2021)}]{Piris2021-xo}%
  \BibitemOpen
  \bibfield  {author} {\bibinfo {author} {\bibfnamefont {M.}~\bibnamefont {Piris}}\ and\ \bibinfo {author} {\bibfnamefont {I.}~\bibnamefont {Mitxelena}},\ } {\bibfield  {journal} {\bibinfo  {journal} {Comput. Phys. Commun.}\ }\textbf {\bibinfo {volume} {259}},\ \bibinfo {pages} {107651} (\bibinfo {year} {2021})}\BibitemShut {NoStop}%
\bibitem [{\citenamefont {Hehre}\ \emph {et~al.}(1969)\citenamefont {Hehre}, \citenamefont {Stewart},\ and\ \citenamefont {Pople}}]{Hehre1969-es}%
  \BibitemOpen
  \bibfield  {author} {\bibinfo {author} {\bibfnamefont {W.~J.}\ \bibnamefont {Hehre}}, \bibinfo {author} {\bibfnamefont {R.~F.}\ \bibnamefont {Stewart}},\ and\ \bibinfo {author} {\bibfnamefont {J.~A.}\ \bibnamefont {Pople}},\ } {\bibfield  {journal} {\bibinfo  {journal} {J. Chem. Phys.}\ }\textbf {\bibinfo {volume} {51}},\ \bibinfo {pages} {2657} (\bibinfo {year} {1969})}\BibitemShut {NoStop}%
\bibitem [{\citenamefont {Dunning}(1989)}]{Dunning1989-mo}%
  \BibitemOpen
  \bibfield  {author} {\bibinfo {author} {\bibfnamefont {T.~H.}\ \bibnamefont {Dunning}, \bibfnamefont {Jr}},\ } {\bibfield  {journal} {\bibinfo  {journal} {J. Chem. Phys.}\ }\textbf {\bibinfo {volume} {90}},\ \bibinfo {pages} {1007} (\bibinfo {year} {1989})}\BibitemShut {NoStop}%
\bibitem [{\citenamefont {Luo}\ \emph {et~al.}(2016)\citenamefont {Luo}, \citenamefont {Benali}, \citenamefont {Shulenburger}, \citenamefont {Krogel}, \citenamefont {Heinonen},\ and\ \citenamefont {Kent}}]{Luo2016-zm}%
  \BibitemOpen
  \bibfield  {author} {\bibinfo {author} {\bibfnamefont {Y.}~\bibnamefont {Luo}}, \bibinfo {author} {\bibfnamefont {A.}~\bibnamefont {Benali}}, \bibinfo {author} {\bibfnamefont {L.}~\bibnamefont {Shulenburger}}, \bibinfo {author} {\bibfnamefont {J.~T.}\ \bibnamefont {Krogel}}, \bibinfo {author} {\bibfnamefont {O.}~\bibnamefont {Heinonen}},\ and\ \bibinfo {author} {\bibfnamefont {P.~R.~C.}\ \bibnamefont {Kent}},\ } {\bibfield  {journal} {\bibinfo  {journal} {New J. Phys.}\ }\textbf {\bibinfo {volume} {18}},\ \bibinfo {pages} {113049} (\bibinfo {year} {2016})}\BibitemShut {NoStop}%
\bibitem [{\citenamefont {Hongo}\ \emph {et~al.}(2015)\citenamefont {Hongo}, \citenamefont {Watson}, \citenamefont {Iitaka}, \citenamefont {Aspuru-Guzik},\ and\ \citenamefont {Maezono}}]{Hongo2015-xt}%
  \BibitemOpen
  \bibfield  {author} {\bibinfo {author} {\bibfnamefont {K.}~\bibnamefont {Hongo}}, \bibinfo {author} {\bibfnamefont {M.~A.}\ \bibnamefont {Watson}}, \bibinfo {author} {\bibfnamefont {T.}~\bibnamefont {Iitaka}}, \bibinfo {author} {\bibfnamefont {A.}~\bibnamefont {Aspuru-Guzik}},\ and\ \bibinfo {author} {\bibfnamefont {R.}~\bibnamefont {Maezono}},\ } {\bibfield  {journal} {\bibinfo  {journal} {J. Chem. Theory Comput.}\ }\textbf {\bibinfo {volume} {11}},\ \bibinfo {pages} {907} (\bibinfo {year} {2015})}\BibitemShut {NoStop}%
\bibitem [{\citenamefont {Scemama}\ \emph {et~al.}(2013)\citenamefont {Scemama}, \citenamefont {Caffarel}, \citenamefont {Oseret},\ and\ \citenamefont {Jalby}}]{Scemama2013-kd}%
  \BibitemOpen
  \bibfield  {author} {\bibinfo {author} {\bibfnamefont {A.}~\bibnamefont {Scemama}}, \bibinfo {author} {\bibfnamefont {M.}~\bibnamefont {Caffarel}}, \bibinfo {author} {\bibfnamefont {E.}~\bibnamefont {Oseret}},\ and\ \bibinfo {author} {\bibfnamefont {W.}~\bibnamefont {Jalby}},\ } {\bibfield  {journal} {\bibinfo  {journal} {J. Comput. Chem.}\ }\textbf {\bibinfo {volume} {34}},\ \bibinfo {pages} {938} (\bibinfo {year} {2013})}\BibitemShut {NoStop}%
\bibitem [{\citenamefont {Mitxelena}\ and\ \citenamefont {Piris}(2022)}]{Mitxelena2022-wt}%
  \BibitemOpen
  \bibfield  {author} {\bibinfo {author} {\bibfnamefont {I.}~\bibnamefont {Mitxelena}}\ and\ \bibinfo {author} {\bibfnamefont {M.}~\bibnamefont {Piris}},\ } {\bibfield  {journal} {\bibinfo  {journal} {J. Chem. Phys.}\ }\textbf {\bibinfo {volume} {156}},\ \bibinfo {pages} {214102} (\bibinfo {year} {2022})}\BibitemShut {NoStop}%
\bibitem [{\citenamefont {Sinitskiy}\ \emph {et~al.}(2010)\citenamefont {Sinitskiy}, \citenamefont {Greenman},\ and\ \citenamefont {Mazziotti}}]{Sinitskiy2010-ik}%
  \BibitemOpen
  \bibfield  {author} {\bibinfo {author} {\bibfnamefont {A.~V.}\ \bibnamefont {Sinitskiy}}, \bibinfo {author} {\bibfnamefont {L.}~\bibnamefont {Greenman}},\ and\ \bibinfo {author} {\bibfnamefont {D.~A.}\ \bibnamefont {Mazziotti}},\ } {\bibfield  {journal} {\bibinfo  {journal} {J. Chem. Phys.}\ }\textbf {\bibinfo {volume} {133}},\ \bibinfo {pages} {014104} (\bibinfo {year} {2010})}\BibitemShut {NoStop}%
\bibitem [{\citenamefont {Mott}(1990)}]{mott1990}%
  \BibitemOpen
  \bibfield  {author} {\bibinfo {author} {\bibfnamefont {N.}~\bibnamefont {Mott}},\ } {\emph {\bibinfo {title} {Metal-Insulator Transitions}}},\ \bibinfo {edition} {2nd}\ ed.\ (\bibinfo  {publisher} {Taylor \& Francis},\ \bibinfo {address} {London},\ \bibinfo {year} {1990})\BibitemShut {NoStop}%
\bibitem [{\citenamefont {Tomanek}\ and\ \citenamefont {Frederick}(2019)}]{TomanekUnknown-hc}%
  \BibitemOpen
  \bibfield  {author} {\bibinfo {author} {\bibfnamefont {D.}~\bibnamefont {Tomanek}}\ and\ \bibinfo {author} {\bibfnamefont {N.}~\bibnamefont {Frederick}},\ } {\bibinfo {title} {Carbon fullerenes}},\ \bibinfo {howpublished} {\url{https://nanotube.msu.edu/fullerene/fullerene-isomers.html}} (\bibinfo {year} {2019}),\ \bibinfo {note} {accessed: 2024-8-2}\BibitemShut {NoStop}%
\bibitem [{\citenamefont {Varganov}\ \emph {et~al.}(2002)\citenamefont {Varganov}, \citenamefont {Avramov}, \citenamefont {Ovchinnikov},\ and\ \citenamefont {Gordon}}]{Varganov2002-dt}%
  \BibitemOpen
  \bibfield  {author} {\bibinfo {author} {\bibfnamefont {S.~A.}\ \bibnamefont {Varganov}}, \bibinfo {author} {\bibfnamefont {P.~V.}\ \bibnamefont {Avramov}}, \bibinfo {author} {\bibfnamefont {S.~G.}\ \bibnamefont {Ovchinnikov}},\ and\ \bibinfo {author} {\bibfnamefont {M.~S.}\ \bibnamefont {Gordon}},\ } {\bibfield  {journal} {\bibinfo  {journal} {Chem. Phys. Lett.}\ }\textbf {\bibinfo {volume} {362}},\ \bibinfo {pages} {380} (\bibinfo {year} {2002})}\BibitemShut {NoStop}%
\bibitem [{\citenamefont {Lee}\ and\ \citenamefont {Head-Gordon}(2019)}]{Lee2019-ju}%
  \BibitemOpen
  \bibfield  {author} {\bibinfo {author} {\bibfnamefont {J.}~\bibnamefont {Lee}}\ and\ \bibinfo {author} {\bibfnamefont {M.}~\bibnamefont {Head-Gordon}},\ }\href {https://doi.org/10.1039/C8CP07613H} {\bibfield  {journal} {\bibinfo  {journal} {Phys. Chem. Chem. Phys.}\ }\textbf {\bibinfo {volume} {21}},\ \bibinfo {pages} {4763} (\bibinfo {year} {2019})}\BibitemShut {NoStop}%
\bibitem [{\citenamefont {Lee}\ \emph {et~al.}(2020)\citenamefont {Lee}, \citenamefont {Malone},\ and\ \citenamefont {Morales}}]{Lee2020-vg}%
  \BibitemOpen
  \bibfield  {author} {\bibinfo {author} {\bibfnamefont {J.}~\bibnamefont {Lee}}, \bibinfo {author} {\bibfnamefont {F.~D.}\ \bibnamefont {Malone}},\ and\ \bibinfo {author} {\bibfnamefont {M.~A.}\ \bibnamefont {Morales}},\ } {\bibfield  {journal} {\bibinfo  {journal} {J. Chem. Theory Comput.}\ }\textbf {\bibinfo {volume} {16}},\ \bibinfo {pages} {3019} (\bibinfo {year} {2020})}\BibitemShut {NoStop}%
\bibitem [{\citenamefont {Mullinax}\ \emph {et~al.}(2019)\citenamefont {Mullinax}, \citenamefont {Epifanovsky}, \citenamefont {Gidofalvi},\ and\ \citenamefont {DePrince}}]{Mullinax2019-km}%
  \BibitemOpen
  \bibfield  {author} {\bibinfo {author} {\bibfnamefont {J.~W.}\ \bibnamefont {Mullinax}}, \bibinfo {author} {\bibfnamefont {E.}~\bibnamefont {Epifanovsky}}, \bibinfo {author} {\bibfnamefont {G.}~\bibnamefont {Gidofalvi}},\ and\ \bibinfo {author} {\bibfnamefont {A.~E.}\ \bibnamefont {DePrince}, \bibfnamefont {3rd}},\ } {\bibfield  {journal} {\bibinfo  {journal} {J. Chem. Theory Comput.}\ }\textbf {\bibinfo {volume} {15}},\ \bibinfo {pages} {276} (\bibinfo {year} {2019})}\BibitemShut {NoStop}%
\bibitem [{\citenamefont {Marian}\ and\ \citenamefont {Gilka}(2008)}]{Marian2008-hz}%
  \BibitemOpen
  \bibfield  {author} {\bibinfo {author} {\bibfnamefont {C.~M.}\ \bibnamefont {Marian}}\ and\ \bibinfo {author} {\bibfnamefont {N.}~\bibnamefont {Gilka}},\ } {\bibfield  {journal} {\bibinfo  {journal} {J. Chem. Theory Comput.}\ }\textbf {\bibinfo {volume} {4}},\ \bibinfo {pages} {1501} (\bibinfo {year} {2008})}\BibitemShut {NoStop}%
\bibitem [{\citenamefont {Grimme}\ and\ \citenamefont {Parac}(2003)}]{Grimme2003-cw}%
  \BibitemOpen
  \bibfield  {author} {\bibinfo {author} {\bibfnamefont {S.}~\bibnamefont {Grimme}}\ and\ \bibinfo {author} {\bibfnamefont {M.}~\bibnamefont {Parac}},\ } {\bibfield  {journal} {\bibinfo  {journal} {ChemPhysChem}\ }\textbf {\bibinfo {volume} {4}},\ \bibinfo {pages} {292} (\bibinfo {year} {2003})}\BibitemShut {NoStop}%
\bibitem [{\citenamefont {Grimme}\ and\ \citenamefont {Waletzke}(1999)}]{Grimme1999-hm}%
  \BibitemOpen
  \bibfield  {author} {\bibinfo {author} {\bibfnamefont {S.}~\bibnamefont {Grimme}}\ and\ \bibinfo {author} {\bibfnamefont {M.}~\bibnamefont {Waletzke}},\ } {\bibfield  {journal} {\bibinfo  {journal} {J. Chem. Phys.}\ }\textbf {\bibinfo {volume} {111}},\ \bibinfo {pages} {5645} (\bibinfo {year} {1999})}\BibitemShut {NoStop}%
\bibitem [{\citenamefont {Angliker}\ \emph {et~al.}(1982)\citenamefont {Angliker}, \citenamefont {Rommel},\ and\ \citenamefont {Wirz}}]{Angliker1982-lq}%
  \BibitemOpen
  \bibfield  {author} {\bibinfo {author} {\bibfnamefont {H.}~\bibnamefont {Angliker}}, \bibinfo {author} {\bibfnamefont {E.}~\bibnamefont {Rommel}},\ and\ \bibinfo {author} {\bibfnamefont {J.}~\bibnamefont {Wirz}},\ } {\bibfield  {journal} {\bibinfo  {journal} {Chem. Phys. Lett.}\ }\textbf {\bibinfo {volume} {87}},\ \bibinfo {pages} {208} (\bibinfo {year} {1982})}\BibitemShut {NoStop}%
\bibitem [{\citenamefont {Burgos}\ \emph {et~al.}(1977)\citenamefont {Burgos}, \citenamefont {Pope}, \citenamefont {Swenberg},\ and\ \citenamefont {Alfano}}]{Burgos1977-ff}%
  \BibitemOpen
  \bibfield  {author} {\bibinfo {author} {\bibfnamefont {J.}~\bibnamefont {Burgos}}, \bibinfo {author} {\bibfnamefont {M.}~\bibnamefont {Pope}}, \bibinfo {author} {\bibfnamefont {C.~E.}\ \bibnamefont {Swenberg}},\ and\ \bibinfo {author} {\bibfnamefont {R.~R.}\ \bibnamefont {Alfano}},\ } {\bibfield  {journal} {\bibinfo  {journal} {Phys. Status Solidi B Basic Res.}\ }\textbf {\bibinfo {volume} {83}},\ \bibinfo {pages} {249} (\bibinfo {year} {1977})}\BibitemShut {NoStop}%
\bibitem [{\citenamefont {Hajgató}\ \emph {et~al.}(2011)\citenamefont {Hajgató}, \citenamefont {Huzak},\ and\ \citenamefont {Deleuze}}]{Hajgato2011-qw}%
  \BibitemOpen
  \bibfield  {author} {\bibinfo {author} {\bibfnamefont {B.}~\bibnamefont {Hajgató}}, \bibinfo {author} {\bibfnamefont {M.}~\bibnamefont {Huzak}},\ and\ \bibinfo {author} {\bibfnamefont {M.~S.}\ \bibnamefont {Deleuze}},\ } {\bibfield  {journal} {\bibinfo  {journal} {J. Phys. Chem. A}\ }\textbf {\bibinfo {volume} {115}},\ \bibinfo {pages} {9282} (\bibinfo {year} {2011})}\BibitemShut {NoStop}%
\bibitem [{\citenamefont {Sony}\ and\ \citenamefont {Shukla}(2007)}]{Sony2007-hu}%
  \BibitemOpen
  \bibfield  {author} {\bibinfo {author} {\bibfnamefont {P.}~\bibnamefont {Sony}}\ and\ \bibinfo {author} {\bibfnamefont {A.}~\bibnamefont {Shukla}},\ } {\bibfield  {journal} {\bibinfo  {journal} {Phys. Rev. B}\ }\textbf {\bibinfo {volume} {75}},\ \bibinfo {pages} {155208} (\bibinfo {year} {2007})}\BibitemShut {NoStop}%
\bibitem [{\citenamefont {Yang}\ \emph {et~al.}(2016)\citenamefont {Yang}, \citenamefont {Davidson},\ and\ \citenamefont {Yang}}]{Yang2016-ql}%
  \BibitemOpen
  \bibfield  {author} {\bibinfo {author} {\bibfnamefont {Y.}~\bibnamefont {Yang}}, \bibinfo {author} {\bibfnamefont {E.~R.}\ \bibnamefont {Davidson}},\ and\ \bibinfo {author} {\bibfnamefont {W.}~\bibnamefont {Yang}},\ } {\bibfield  {journal} {\bibinfo  {journal} {Proc. Natl. Acad. Sci. U. S. A.}\ }\textbf {\bibinfo {volume} {113}},\ \bibinfo {pages} {E5098} (\bibinfo {year} {2016})}\BibitemShut {NoStop}%
\bibitem [{\citenamefont {Dupuy}\ and\ \citenamefont {Casula}(2018)}]{Dupuy2018-nc}%
  \BibitemOpen
  \bibfield  {author} {\bibinfo {author} {\bibfnamefont {N.}~\bibnamefont {Dupuy}}\ and\ \bibinfo {author} {\bibfnamefont {M.}~\bibnamefont {Casula}},\ } {\bibfield  {journal} {\bibinfo  {journal} {J. Chem. Phys.}\ }\textbf {\bibinfo {volume} {148}} (\bibinfo {year} {2018})}\BibitemShut {NoStop}%
\bibitem [{\citenamefont {Zhuang}\ \emph {et~al.}(2020)\citenamefont {Zhuang}, \citenamefont {Tang}, \citenamefont {Ding}, \citenamefont {Tatikonda}, \citenamefont {Dvornek}, \citenamefont {Papademetris},\ and\ \citenamefont {Duncan}}]{Zhuang2020-of}%
  \BibitemOpen
  \bibfield  {author} {\bibinfo {author} {\bibfnamefont {J.}~\bibnamefont {Zhuang}}, \bibinfo {author} {\bibfnamefont {T.}~\bibnamefont {Tang}}, \bibinfo {author} {\bibfnamefont {Y.}~\bibnamefont {Ding}}, \bibinfo {author} {\bibfnamefont {S.}~\bibnamefont {Tatikonda}}, \bibinfo {author} {\bibfnamefont {N.}~\bibnamefont {Dvornek}}, \bibinfo {author} {\bibfnamefont {X.}~\bibnamefont {Papademetris}},\ and\ \bibinfo {author} {\bibfnamefont {J.~S.}\ \bibnamefont {Duncan}},\ } {\bibfield  {journal} {\bibinfo  {journal} {arXiv [cs.LG]}\ } (\bibinfo {year} {2020})}\BibitemShut {NoStop}%
\bibitem [{\citenamefont {Zaheer}\ \emph {et~al.}(2018)\citenamefont {Zaheer}, \citenamefont {Reddi}, \citenamefont {Sachan}, \citenamefont {Kale},\ and\ \citenamefont {Kumar}}]{Zaheer2018-ek}%
  \BibitemOpen
  \bibfield  {author} {\bibinfo {author} {\bibfnamefont {M.}~\bibnamefont {Zaheer}}, \bibinfo {author} {\bibfnamefont {S.~J.}\ \bibnamefont {Reddi}}, \bibinfo {author} {\bibfnamefont {D.}~\bibnamefont {Sachan}}, \bibinfo {author} {\bibfnamefont {S.}~\bibnamefont {Kale}},\ and\ \bibinfo {author} {\bibfnamefont {S.}~\bibnamefont {Kumar}},\ }in\  {\emph {\bibinfo {booktitle} {Proceedings of the 32nd International Conference on Neural Information Processing Systems}}},\ \bibinfo {series and number} {NIPS'18}\ (\bibinfo  {publisher} {Curran Associates Inc.},\ \bibinfo {address} {Red Hook, NY, USA},\ \bibinfo {year} {2018})\ pp.\ \bibinfo {pages} {9815--9825}\BibitemShut {NoStop}%
\bibitem [{\citenamefont {Schmidt}\ \emph {et~al.}(2019)\citenamefont {Schmidt}, \citenamefont {Benavides-Riveros},\ and\ \citenamefont {Marques}}]{Schmidt2019-gs}%
  \BibitemOpen
  \bibfield  {author} {\bibinfo {author} {\bibfnamefont {J.}~\bibnamefont {Schmidt}}, \bibinfo {author} {\bibfnamefont {C.~L.}\ \bibnamefont {Benavides-Riveros}},\ and\ \bibinfo {author} {\bibfnamefont {M.~A.~L.}\ \bibnamefont {Marques}},\ } {\bibfield  {journal} {\bibinfo  {journal} {Phys. Rev. B}\ }\textbf {\bibinfo {volume} {99}},\ \bibinfo {pages} {224502} (\bibinfo {year} {2019})}\BibitemShut {NoStop}%
\end{thebibliography}

\begin{thebibliography}{6}%
\makeatletter
\providecommand \@ifxundefined [1]{%
 \@ifx{#1\undefined}
}%
\providecommand \@ifnum [1]{%
 \ifnum #1\expandafter \@firstoftwo
 \else \expandafter \@secondoftwo
 \fi
}%
\providecommand \@ifx [1]{%
 \ifx #1\expandafter \@firstoftwo
 \else \expandafter \@secondoftwo
 \fi
}%
\providecommand \natexlab [1]{#1}%
\providecommand \enquote  [1]{``#1''}%
\providecommand \bibnamefont  [1]{#1}%
\providecommand \bibfnamefont [1]{#1}%
\providecommand \citenamefont [1]{#1}%
\providecommand \@href[1]{\@@startlink{#1}\@@href}%
\providecommand \@@href[1]{\endgroup#1\@@endlink}%
\providecommand \@sanitize@url [0]{\catcode `\\12\catcode `\$12\catcode `\&12\catcode `\#12\catcode `\^12\catcode `\_12\catcode `\%12\relax}%
\providecommand \@@startlink[1]{}%
\providecommand \@@endlink[0]{}%
\providecommand \@url [1]{\endgroup\@href {#1}{\urlprefix }}%
\providecommand \urlprefix  [0]{URL }%
\providecommand \doibase [0]{https://doi.org/}%
\providecommand \selectlanguage [0]{\@gobble}%
\providecommand \bibinfo  [0]{\@secondoftwo}%
\providecommand \bibfield  [0]{\@secondoftwo}%
\providecommand \translation [1]{[#1]}%
\providecommand \BibitemOpen [0]{}%
\providecommand \bibitemStop [0]{}%
\providecommand \bibitemNoStop [0]{.\EOS\space}%
\providecommand \EOS [0]{\spacefactor3000\relax}%
\providecommand \BibitemShut  [1]{\csname bibitem#1\endcsname}%
\let\auto@bib@innerbib\@empty
%</preamble>
\bibitem [{\citenamefont {Paszke}\ \emph {et~al.}(2019)\citenamefont {Paszke}, \citenamefont {Gross}, \citenamefont {Massa}, \citenamefont {Lerer}, \citenamefont {Bradbury}, \citenamefont {Chanan}, \citenamefont {Killeen}, \citenamefont {Lin}, \citenamefont {Gimelshein}, \citenamefont {Antiga}, \citenamefont {Desmaison}, \citenamefont {Köpf}, \citenamefont {Yang}, \citenamefont {DeVito}, \citenamefont {Raison}, \citenamefont {Tejani}, \citenamefont {Chilamkurthy}, \citenamefont {Steiner}, \citenamefont {Fang}, \citenamefont {Bai},\ and\ \citenamefont {Chintala}}]{Paszke2019-fu}%
  \BibitemOpen
  \bibfield  {author} {\bibinfo {author} {\bibfnamefont {A.}~\bibnamefont {Paszke}}, \bibinfo {author} {\bibfnamefont {S.}~\bibnamefont {Gross}}, \bibinfo {author} {\bibfnamefont {F.}~\bibnamefont {Massa}}, \bibinfo {author} {\bibfnamefont {A.}~\bibnamefont {Lerer}}, \bibinfo {author} {\bibfnamefont {J.}~\bibnamefont {Bradbury}}, \bibinfo {author} {\bibfnamefont {G.}~\bibnamefont {Chanan}}, \bibinfo {author} {\bibfnamefont {T.}~\bibnamefont {Killeen}}, \bibinfo {author} {\bibfnamefont {Z.}~\bibnamefont {Lin}}, \bibinfo {author} {\bibfnamefont {N.}~\bibnamefont {Gimelshein}}, \bibinfo {author} {\bibfnamefont {L.}~\bibnamefont {Antiga}}, \bibinfo {author} {\bibfnamefont {A.}~\bibnamefont {Desmaison}}, \bibinfo {author} {\bibfnamefont {A.}~\bibnamefont {Köpf}}, \bibinfo {author} {\bibfnamefont {E.}~\bibnamefont {Yang}}, \bibinfo {author} {\bibfnamefont {Z.}~\bibnamefont {DeVito}}, \bibinfo {author} {\bibfnamefont {M.}~\bibnamefont {Raison}}, \bibinfo {author} {\bibfnamefont {A.}~\bibnamefont {Tejani}}, \bibinfo
  {author} {\bibfnamefont {S.}~\bibnamefont {Chilamkurthy}}, \bibinfo {author} {\bibfnamefont {B.}~\bibnamefont {Steiner}}, \bibinfo {author} {\bibfnamefont {L.}~\bibnamefont {Fang}}, \bibinfo {author} {\bibfnamefont {J.}~\bibnamefont {Bai}},\ and\ \bibinfo {author} {\bibfnamefont {S.}~\bibnamefont {Chintala}},\ } {\bibfield  {journal} {\bibinfo  {journal} {arXiv [cs.LG]}\ } (\bibinfo {year} {2019})}\BibitemShut {NoStop}%
\bibitem [{\citenamefont {DeepMind}\ \emph {et~al.}(2020)\citenamefont {DeepMind}, \citenamefont {Babuschkin}, \citenamefont {Baumli}, \citenamefont {Bell}, \citenamefont {Bhupatiraju}, \citenamefont {Bruce}, \citenamefont {Buchlovsky}, \citenamefont {Budden}, \citenamefont {Cai}, \citenamefont {Clark}, \citenamefont {Danihelka}, \citenamefont {Dedieu}, \citenamefont {Fantacci}, \citenamefont {Godwin}, \citenamefont {Jones}, \citenamefont {Hemsley}, \citenamefont {Hennigan}, \citenamefont {Hessel}, \citenamefont {Hou}, \citenamefont {Kapturowski}, \citenamefont {Keck}, \citenamefont {Kemaev}, \citenamefont {King}, \citenamefont {Kunesch}, \citenamefont {Martens}, \citenamefont {Merzic}, \citenamefont {Mikulik}, \citenamefont {Norman}, \citenamefont {Papamakarios}, \citenamefont {Quan}, \citenamefont {Ring}, \citenamefont {Ruiz}, \citenamefont {Sanchez}, \citenamefont {Sartran}, \citenamefont {Schneider}, \citenamefont {Sezener}, \citenamefont {Spencer}, \citenamefont {Srinivasan}, \citenamefont
  {Stanojevi\'{c}}, \citenamefont {Stokowiec}, \citenamefont {Wang}, \citenamefont {Zhou},\ and\ \citenamefont {Viola}}]{deepmind2020jax}%
  \BibitemOpen
  \bibfield  {author} {\bibinfo {author} {\bibnamefont {DeepMind}}, \bibinfo {author} {\bibfnamefont {I.}~\bibnamefont {Babuschkin}}, \bibinfo {author} {\bibfnamefont {K.}~\bibnamefont {Baumli}}, \bibinfo {author} {\bibfnamefont {A.}~\bibnamefont {Bell}}, \bibinfo {author} {\bibfnamefont {S.}~\bibnamefont {Bhupatiraju}}, \bibinfo {author} {\bibfnamefont {J.}~\bibnamefont {Bruce}}, \bibinfo {author} {\bibfnamefont {P.}~\bibnamefont {Buchlovsky}}, \bibinfo {author} {\bibfnamefont {D.}~\bibnamefont {Budden}}, \bibinfo {author} {\bibfnamefont {T.}~\bibnamefont {Cai}}, \bibinfo {author} {\bibfnamefont {A.}~\bibnamefont {Clark}}, \bibinfo {author} {\bibfnamefont {I.}~\bibnamefont {Danihelka}}, \bibinfo {author} {\bibfnamefont {A.}~\bibnamefont {Dedieu}}, \bibinfo {author} {\bibfnamefont {C.}~\bibnamefont {Fantacci}}, \bibinfo {author} {\bibfnamefont {J.}~\bibnamefont {Godwin}}, \bibinfo {author} {\bibfnamefont {C.}~\bibnamefont {Jones}}, \bibinfo {author} {\bibfnamefont {R.}~\bibnamefont {Hemsley}}, \bibinfo
  {author} {\bibfnamefont {T.}~\bibnamefont {Hennigan}}, \bibinfo {author} {\bibfnamefont {M.}~\bibnamefont {Hessel}}, \bibinfo {author} {\bibfnamefont {S.}~\bibnamefont {Hou}}, \bibinfo {author} {\bibfnamefont {S.}~\bibnamefont {Kapturowski}}, \bibinfo {author} {\bibfnamefont {T.}~\bibnamefont {Keck}}, \bibinfo {author} {\bibfnamefont {I.}~\bibnamefont {Kemaev}}, \bibinfo {author} {\bibfnamefont {M.}~\bibnamefont {King}}, \bibinfo {author} {\bibfnamefont {M.}~\bibnamefont {Kunesch}}, \bibinfo {author} {\bibfnamefont {L.}~\bibnamefont {Martens}}, \bibinfo {author} {\bibfnamefont {H.}~\bibnamefont {Merzic}}, \bibinfo {author} {\bibfnamefont {V.}~\bibnamefont {Mikulik}}, \bibinfo {author} {\bibfnamefont {T.}~\bibnamefont {Norman}}, \bibinfo {author} {\bibfnamefont {G.}~\bibnamefont {Papamakarios}}, \bibinfo {author} {\bibfnamefont {J.}~\bibnamefont {Quan}}, \bibinfo {author} {\bibfnamefont {R.}~\bibnamefont {Ring}}, \bibinfo {author} {\bibfnamefont {F.}~\bibnamefont {Ruiz}}, \bibinfo {author} {\bibfnamefont
  {A.}~\bibnamefont {Sanchez}}, \bibinfo {author} {\bibfnamefont {L.}~\bibnamefont {Sartran}}, \bibinfo {author} {\bibfnamefont {R.}~\bibnamefont {Schneider}}, \bibinfo {author} {\bibfnamefont {E.}~\bibnamefont {Sezener}}, \bibinfo {author} {\bibfnamefont {S.}~\bibnamefont {Spencer}}, \bibinfo {author} {\bibfnamefont {S.}~\bibnamefont {Srinivasan}}, \bibinfo {author} {\bibfnamefont {M.}~\bibnamefont {Stanojevi\'{c}}}, \bibinfo {author} {\bibfnamefont {W.}~\bibnamefont {Stokowiec}}, \bibinfo {author} {\bibfnamefont {L.}~\bibnamefont {Wang}}, \bibinfo {author} {\bibfnamefont {G.}~\bibnamefont {Zhou}},\ and\ \bibinfo {author} {\bibfnamefont {F.}~\bibnamefont {Viola}},\ } {\bibinfo {title} {The {D}eep{M}ind {JAX} {E}cosystem}} (\bibinfo {year} {2020}),\ \bibinfo {note} {http://github.com/google-deepmind}\BibitemShut {NoStop}%
\bibitem [{\citenamefont {Kingma}\ and\ \citenamefont {Ba}(2014)}]{Kingma2014-xg}%
  \BibitemOpen
  \bibfield  {author} {\bibinfo {author} {\bibfnamefont {D.~P.}\ \bibnamefont {Kingma}}\ and\ \bibinfo {author} {\bibfnamefont {J.}~\bibnamefont {Ba}},\ } {\bibfield  {journal} {\bibinfo  {journal} {arXiv:1412.6980 [cs.LG]}\ } (\bibinfo {year} {2014})}\BibitemShut {NoStop}%
\bibitem [{\citenamefont {Franco}\ \emph {et~al.}(2024)\citenamefont {Franco}, \citenamefont {Bonfil-Rivera}, \citenamefont {Huan Lew-Yee}, \citenamefont {Piris}, \citenamefont {M~Del~Campo},\ and\ \citenamefont {Vargas-Hernández}}]{Franco2024-sh}%
  \BibitemOpen
  \bibfield  {author} {\bibinfo {author} {\bibfnamefont {L.}~\bibnamefont {Franco}}, \bibinfo {author} {\bibfnamefont {I.~A.}\ \bibnamefont {Bonfil-Rivera}}, \bibinfo {author} {\bibfnamefont {J.~F.}\ \bibnamefont {Huan Lew-Yee}}, \bibinfo {author} {\bibfnamefont {M.}~\bibnamefont {Piris}}, \bibinfo {author} {\bibfnamefont {J.}~\bibnamefont {M~Del~Campo}},\ and\ \bibinfo {author} {\bibfnamefont {R.~A.}\ \bibnamefont {Vargas-Hernández}},\ } {\bibfield  {journal} {\bibinfo  {journal} {J. Chem. Phys.}\ }\textbf {\bibinfo {volume} {160}},\ \bibinfo {pages} {244107} (\bibinfo {year} {2024})}\BibitemShut {NoStop}%
\bibitem [{\citenamefont {Piris}(2021)}]{Piris2021-sv}%
  \BibitemOpen
  \bibfield  {author} {\bibinfo {author} {\bibfnamefont {M.}~\bibnamefont {Piris}},\ } {\bibfield  {journal} {\bibinfo  {journal} {Phys. Rev. Lett.}\ }\textbf {\bibinfo {volume} {127}},\ \bibinfo {pages} {233001} (\bibinfo {year} {2021})}\BibitemShut {NoStop}%
\bibitem [{\citenamefont {Pritchard}\ \emph {et~al.}(2019)\citenamefont {Pritchard}, \citenamefont {Altarawy}, \citenamefont {Didier}, \citenamefont {Gibson},\ and\ \citenamefont {Windus}}]{BSE2019}%
  \BibitemOpen
  \bibfield  {author} {\bibinfo {author} {\bibfnamefont {B.~P.}\ \bibnamefont {Pritchard}}, \bibinfo {author} {\bibfnamefont {D.}~\bibnamefont {Altarawy}}, \bibinfo {author} {\bibfnamefont {B.}~\bibnamefont {Didier}}, \bibinfo {author} {\bibfnamefont {T.~D.}\ \bibnamefont {Gibson}},\ and\ \bibinfo {author} {\bibfnamefont {T.~L.}\ \bibnamefont {Windus}},\ }\href {https://doi.org/10.1021/acs.jcim.9b00725} {\bibfield  {journal} {\bibinfo  {journal} {J. Chem. Inf. Model.}\ }\textbf {\bibinfo {volume} {59}},\ \bibinfo {pages} {4814 } (\bibinfo {year} {2019})}\BibitemShut {NoStop}%
\end{thebibliography}

\providecommand{\noopsort}[1]{}\providecommand{\singleletter}[1]{#1}%

\section{Supporting Information}

\subsection{Algorithm Details}

Here, we present the algorithm used in this work for optimizing natural orbitals (NOs) and occupation numbers (ONs) within the framework of Natural Orbital Functional Theory (NOFT). The optimization follows a two-step procedure in which NOs and ONs are optimized separately, each with its own set of internal iterations. Globally, an NO optimization cycle is performed with fixed occupations, followed by an ON optimization cycle with fixed orbitals. Together, these steps constitute an external iteration. This process is repeated until convergence is achieved, as illustrated in Algorithm~\ref{alg:algorithm}.

A key factor for achieving convergence is the use of the adaptive momentum (ADAM) method to optimize NOs, while ONs are optimized using the conjugate gradient method. Since the overall procedure is deeply inspired by techniques from the field of deep learning, it is useful to highlight some parallels. For instance, an external iteration in NOFT is analogous to an epoch in deep learning.

\begin{algorithm}[H]
  \caption{NOF optimization algortithm.}\label{alg:algorithm}
  \label{Optimization-Algorithm}
   \begin{algorithmic}[1]
\State $\alpha = 0.01$
\While{External Iteration not converged}
\Require Orbital Optimization (ADAM)
\State $\mathbf{m}_0 = \mathbf{0}$, $\mathbf{v}_0 = \mathbf{0}$, $\hat{\mathbf{v}}^{\max}_0 = \mathbf{0}$, $\beta_1=0.7$, $\beta_2=0.9$, $p=0.4$
\For{t in 1:$N_\text{it}^\text{orb}$}
\State Compute $\lambda_{pq}$ (requires $\mathbf{J}$ and $\mathbf{K}$ integrals)
\State $g_{pq} = {\lambda_{pq}-\lambda_{qp}}$;  $\left\{p<q\right\}$
%\State $\alpha=\text{Norm}_\infty(g_{pq})$
\State $\mathbf{m}_t = \beta_1 \mathbf{m}_{t-1} + (1-\beta_1) \mathbf{g}_t$
\State $\mathbf{v}_t = \beta_2 \mathbf{v}_{t-1} + (1-\beta_2) \mathbf{g}^2_t$
\State $\mathbf{\hat{m}}_t = \mathbf{m}_t/(1-\beta_1^t)$
\State $\mathbf{\hat{v}}_t = \mathbf{v}_t/(1-\beta_2^t)$
\State $\mathbf{\hat{v}}_t^{\max} = \max(\hat{\mathbf{v}}_{t-1}^{\max},\hat{\mathbf{v}})$
\State $\mathbf{y}_t = \alpha \hat{\mathbf{m}}_t / \sqrt{\hat{\mathbf{v}}_t^{\max} + \varepsilon}$
\State $\mathbf{U}_t = e^{\mathbf{y}_t}$
\State $\mathbf{C}_t = \mathbf{C}_{t-1} \mathbf{U}_t$
\State Compute $E$
\EndFor
\If{$E$ not improved}
    \State $\alpha=p\alpha$
    \State $N_\text{it}^\text{orb} += 10$
\EndIf
\Require Occupation Number Optimization (CG)
\State Compute ($\mathbf{J}$ and $\mathbf{K}$) integrals
\State Optimize ON under parameterization
\If{ONs were already converged}
    \State $N_\text{it}^\text{orb} += 10$
\EndIf
\EndWhile
\end{algorithmic}
\end{algorithm}

The process starts with the calculation of the orbital gradient. Using the orbital rotation method, the unitary matrix is parameterized as $\mathbf{U}=e^{\mathbf{y}}$. Since $y_{pq}$ are the elements of an antisymmetric matrix ($y_{pq} = - y_{qp}$), only the $N_{bf}(N_{bf}-1)/2$ elements remain free to move. By choosing the vectorization of the upper triangular elements ($p<q$), we can compute the gradient elements as
\begin{equation}
    g_{pq} = \frac{dE}{dy_{pq}} \bigg\rvert_{\mathbf{y}=\mathbf{0}} = 4(\lambda_{pq}-\lambda_{qp}); \>\>  p<q
\end{equation}
with $\mathbf{\lambda}$ the matrix of Lagrange multipliers that enforces the orthonormality of the orbitals; in the case of JK(L) NOFs, these depend on the Coulomb and exchange (and time-exchange) integrals.

The ADAM technique relies on the first ($\mathbf{m}$) and second ($\mathbf{v}$) moments of the gradient, related to the mean and variance, which aid in determining the correct step direction and scaling the step size for each parameter. The element-wise moments at the $t$-iteration are computed as
\begin{eqnarray}
\mathbf{m}_t &=& \beta_1 \mathbf{m}_{t-1} + (1-\beta_1) \mathbf{g}_t\\
\mathbf{v}_t &=& \beta_2 \mathbf{v}_{t-1} + (1-\beta_2) \mathbf{g}^2_t  
\end{eqnarray}
with $\beta_1$ and $\beta_2$ being two parameters that control the influence of the moments of past iterations; boldface notation is used to emphasize compound variables. A bias arises due to the initialization of moments, so a correction is applied as
\begin{eqnarray}
    \hat{\mathbf{m}}_t &=& \mathbf{m}_t/(1-\beta_1^t)\\
    \hat{\mathbf{v}}_t &=& \mathbf{v}_t/(1-\beta_2^t)
\end{eqnarray}

In the deep learning context, taking the maximum of the second momentum through iterations (known as AMSgrad) might improve convergency on some problems. In the case of the natural orbital optimization at hand, we have found it crucial for enhancing convergence

\begin{equation}
    \hat{\mathbf{v}}_t^{\max} = \max(\hat{\mathbf{v}}_{t-1}^{\max},\hat{\mathbf{v}}_t)   
\end{equation}
in this way, the variable $\hat{\mathbf{v}}_t$ continues to evolve through iterations, but only the historical maximum is stored in $\hat{\mathbf{v}}_t^{\max}$.

Finally, the orbital rotation step is given as
\begin{eqnarray}
    \mathbf{y}_t &=& \alpha \hat{\mathbf{m}}_t / \sqrt{\hat{\mathbf{v}}_{\max} + \varepsilon} \label{eq:step}\\
    \mathbf{U}_t &=& e^\mathbf{y_t} \label{eq:U}\\
    \mathbf{C}_t &=& \mathbf{C}_{t-1} \mathbf{U}_t
\end{eqnarray}
with $\alpha$ representing the step length or learning rate and $\varepsilon=10^{-16}$ serving as a small value to prevent the denominator from vanishing. These new orbitals are used to compute the energy and gradients. It is important to note that this step is one of the main differences compared to the conventional implementation of ADAM, since the starting point in Eq.~(\ref{eq:step}) is always $\mathbf{y}_{t-1} = \mathbf{0}$. We also point out that Eq.~(\ref{eq:U}) implies using the vector $\mathbf{y}_t$ to build the antisymmetric matrix required in the exponential. The process is repeated until a breaking condition is achieved for the internal orbital optimization, either by convergence or by reaching the $N_\text{it}^\text{orb}$ iterations (early stopping).

The implementation of the ADAM optimizer used in this work is based on those reported in PyTorch\cite{Paszke2019-fu} and Optax\cite{deepmind2020jax}, which report the default parameter values for the first and second moment estimates set as $\beta_1 = 0.9$ and $\beta_2 = 0.999$, and a learning rate of $\alpha = 0.001$, as proposed in the original paper.\cite{Kingma2014-xg} While convergence can be achieved with these parameters, it typically requires a large number of iterations. We have found better results by adjusting these values, specifically setting $\beta_1 = 0.7$ and $\beta_2 = 0.9$.

Additionally, we observed that convergence is further improved by starting with a higher learning rate (e.g., $\alpha = 0.01$) and gradually reducing it throughout the optimization process (e.g., to $\alpha = 0.001$). This approach mirrors the use of scheduled learning rates in deep learning, where the learning rate is adjusted as a function of the number of iterations (or epochs). In our case, we perform a stepwise reduction by multiplying $\alpha$ by a factor (e.g., $p = 0.4$) whenever the energy fails to improve in a NO optimization cycle. Alternatively, the learning rate can be dynamically adjusted by setting $\alpha$ proportional to the infinite norm of the gradient at the start of each external iteration.

On the other hand, the optimization of the ONs is carried out using the recently proposed softmax parameterization,\cite{Franco2024-sh} which enhances stability, though other parameterizations could also be employed. This parameterization directly incorporates the $\mathbb{N}$-representability conditions, allowing us to perform unconstrained optimization using the conjugate gradient method without encountering additional issues.

An additional and crucial consideration must be given to the control of the optimization procedure. In principle, one could allow the NO and ON optimizations to iterate independently until internal convergence is achieved, which would eventually lead to external convergence. However, this approach may not be the most efficient. This is particularly relevant during the early stages of external iterations, when NO optimization occurs with fixed, sub-optimal ONs. In this case, a large change in the ONs might invalidate previously converged orbitals of the previous step.

Since ON optimization iterations are computationally cheaper than NO optimization iterations, we apply early stopping to NOs optimization and allow ONs optimization to continue until convergence. Additionally, we start with a relatively low number of iterations for the orbital optimization (e.g., $N_\text{it}^\text{orb} = 10$) and progressively increase this number whenever either the NO or ON optimization does not improve the energy. In this way, we couple the optimizations of NOs and ONs while allowing them to be performed separately, and the method invests in the expensive iterations only when they are valuable for convergence.

\begin{figure}[htb]
  \centering
  \begin{tikzpicture}
  \begin{axis}[
      xlabel={External Iteration},
      ylabel={Number of Internal Iterations},
      legend pos=north west,
      grid=major,
    ]
    \addplot[
      %scatter,
      only marks,
      mark=square*,
      mark options={blue},
    ]
    coordinates {
(1, 10)
(2, 10)
(3, 10)
(4, 10)
(5, 10)
(6, 10) 
(7, 10) 
(8, 10) 
(9, 10) 
(10,10) 
(11,10)
(12,10) 
(13,10) 
(14,10) 
(15,10) 
(16,10) 
(17,10) 
(18,10) 
(19,40) 
(20,39) 
(21,38) 
(22,40) 
(23,40) 
};
  \addlegendentry{Natural Orbitals}
    \addplot[
      %scatter,
      only marks,
      mark=square*,
      mark options={orange},
    ]
    coordinates {
(1, 26)
(2, 45)
(3, 21)
(4, 18)
(5, 12)
(6, 12) 
(7, 12) 
(8,  9) 
(9, 12) 
(10, 9) 
(11,10)
(12, 7) 
(13, 5) 
(14, 5) 
(15, 5) 
(16, 7) 
(17, 7) 
(18, 0) 
(19, 6) 
(20, 9) 
(21,10)
(22,10)
(23,12)
};
  \addlegendentry{Occupation Numbers}
  \end{axis}
\end{tikzpicture}
  \caption{Evolution of the internal iterations of natural orbitals and occupation numbers along the external iterations in the optimization of benzene with GNOF/cc-pVDZ.}
  \label{fig:convergency}
\end{figure}
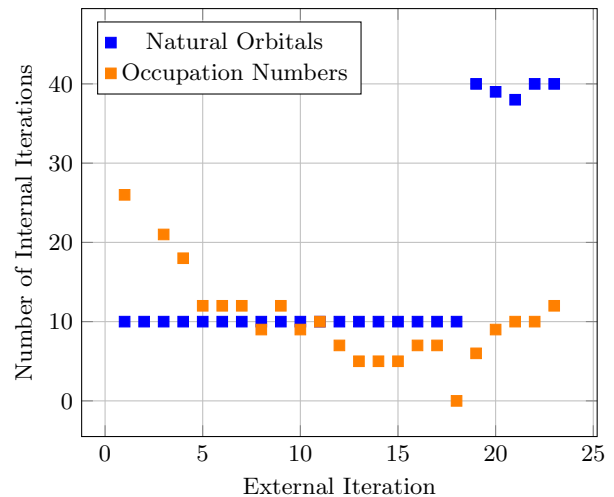

An example of the optimization process is shown in Fig.~\ref{fig:convergency} for a GNOF calculation of benzene. The vertical axis displays the number of internal iterations performed by the NO optimization (blue) and the ON optimization (orange) during each external iteration. At the first external iterations, the optimization is dominated by the cheaper ON optimization, while the more expensive NO optimization is kept at a constant low number of internal iterations due to early stopping. Once the ONs remain unchanged (which happens at the 18$^\text{th}$ external iteration in this example), it triggers an increase in the number of iterations allowed for the NO optimization, enabling the calculation to efficiently reach convergence.

\subsection{GNOF modified (GNOFm)}

The original Global Natural Orbital Functional (GNOF) has been presented in Ref. [5], and correspond to a pairing model, implying the following considerations:
\begin{itemize}
    \item Given a system of $N$ electrons, we divide them into $N=N_\text{I}+N_\text{II}$, with $N_\text{I}$ determined by the spin multiplicity of the system, $2S+1 = 2N_\text{I}+1$ such that $\expval{\hat{S}^2} = S(S+1)$.
    \item Similarly, given $N_\text{orb}$ orbitals in orbital space $\Omega$, they are divided into sets such as $\Omega = \Omega_\text{I} \oplus \Omega_\text{II}$, with the subindex labeling the occupancy of the orbitals in the set, that is, orbitals in $\Omega_\text{II}$ are double occupied and orbitals in $\Omega_\text{I}$ are single occupied.
    \item $\Omega_\text{I}$ is integrated by $N_I$ subspaces $\Omega_g$, each containing one orbital, with occupancy $2n_g=1.0$.
    \item $\Omega_\text{II}$ is integrated by $N_\text{II}/2$ subspaces $\Omega_g$, each with a strongly occupied orbital coupled to $N_\text{cwo}$ orbitals, such that the sum of the occupation numbers in each subspace is $2\sum_{i \in \Omega_g} n_i=2.0$.
    \item Overall, these considerations lead to $N_\Omega$ subspaces. Notable, for singlets, the $\Omega_\text{I}$ is empty since $N_I=0$.
\end{itemize} 

The energy of GNOF is given by:
\begin{equation}
    E = E^\text{intra} + E_\text{HF}^\text{inter} + E_\text{sta}^\text{inter} + E_\text{dyn}^\text{inter}
\end{equation}
with the intrapair energy providing the interactions inside a given subspace $\Omega_g$
\begin{equation}
    E^\text{intra} = \sum_{g=1}^{N_\text{II}/2} E_g + \sum_{g=N_\text{II}/2+1}^{N_\Omega} H_{gg}
\end{equation}
\begin{equation}
    E_g = \sum_{p \in \Omega_g} n_p (2H_{pp}+J_{pp}) + \sum_{q,p \in \Omega_g,  p \neq q} \Pi (n_q, n_p) L_{pq}
\end{equation}
\begin{equation}
    \Pi(n_q,n_p) = \sqrt{n_q n_p} (\delta_{q\Omega_a}\delta_{p\Omega_a} - \delta_{qg} - \delta_{pg})
\end{equation}
with $\delta_{q\Omega_a}$ being 1.0 if $q$ is in an orbital above $N_{\Omega}$ and zero otherwise.

In contrast, the interpair electron correlation provides interactions between orbitals in distinct subspaces. It presents a Hartree-Fock like part
 \begin{equation}
    E_\text{HF}^\text{inter} = \sideset{}{'}\sum_{p,q=1}^{N_\text{orb}} n_q n_p (2 J_{pq} - K_{pq})
\end{equation}
where the prime indicates that the sum omits terms corresponding to $p$ and $q$ in the same subspace.

A static interpar contribution is considered as
\begin{multline}
    E_\text{sta}^\text{inter} = -\bigg( \sum_{p=1}^{N_\Omega}\sum_{q=N_\Omega+1}^{N_\text{orb}}\\ +  \sum_{p=N_\Omega+1}^{N_\text{orb}}\sum_{q=1}^{N_\Omega} + \sum_{p,q=N_\Omega+1}^{N_\text{orb}} \bigg)' \Phi_q \Phi_p L_{pq} \\
    -\frac{1}{2}\bigg( \sum_{p=1}^{N_\text{II}/2}\sum_{q=N_\text{II}/2+1}^{N_\Omega} + \sum_{p=N_\text{II}/2+1}^{N_\Omega}\sum_{q=1}^
    {N_\text{II}/2} \bigg)' \Phi_q \Phi_p L_{pq} \\
    - \sideset{}{'}\sum_{p,q=N_\text{II}/2+1}^{N_\Omega} \Phi_q \Phi_p K_{pq}
\end{multline}
\begin{equation}
    \phi_p = \sqrt{n_p(1-n_p)}
\end{equation}
and a dynamic interpair contribution
\begin{multline}
    E_\text{dyn}^\text{inter} = - \bigg( \sum_{p=1}^{N_\text{II}/2}\sum_{q=N_\Omega+1}^{N_\text{orb}} + \sum_{p=N_\Omega+1}^{N_\text{orb}}\sum_{q=1}^{N_\text{II}/2} \bigg)' \\
    \times \bigg(\sqrt{n_q^d n_p^d} - n_q^d n_p^d\bigg) L_{pq} \\
    + \sideset{}{'}\sum_{p,q=N_\Omega+1}^{N_\text{orb}} \bigg(\sqrt{n_q^d n_p^d} + n_q^d n_p^d\bigg) L_{pq}
\end{multline}
with dynamic occupation numbers computed as
\begin{equation}
    n_p^d = n_p \frac{h_g^d}{h_g}, \>\> p\in \Omega_g, \>\> g=1,2,\cdots,N_\text{II}/2
\end{equation}
being the hole $h_g = 1-n_g$ and the dynamic hole of the strongly correlated orbital
\begin{equation}
    h_g^d = h_g e^{-\left(\frac{h_g}{h_c}\right)^2} \>\> g=1,2,\cdots,N_\text{II}/2
\end{equation}
and $h_c = 0.02\sqrt{2}$.

In the case of the modified GNOF (GNOFm) used in this work, we have adapted the interpair static contribution to be  
\begin{multline}
    E_\text{sta}^\text{inter} = -\bigg(\textcolor{blue}{\sum_{p=1}^{N_\text{II}/2}\sum_{q=1}^{N_\text{II}/2}} + \sum_{p=1}^{N_\Omega}\sum_{q=N_\Omega+1}^{N_\text{orb}}\\ + \sum_{p=N_\Omega+1}^{N_\text{orb}}\sum_{q=1}^{N_\Omega} + \sum_{p,q=N_\Omega+1}^{N_\text{orb}}\bigg)' \Phi_q \Phi_p L_{pq} \\
    -\textcolor{blue}{1}\bigg(\sum_{p=1}^{N_\text{II}/2}\sum_{q=N_\text{II}/2+1}^{N_\Omega} + \sum_{p=N_\text{II}/2+1}^{N_\Omega}\sum_{q=1}^{N_\text{II}/2} \bigg)' \Phi_q \Phi_p L_{pq} \\
    - \sideset{}{'} \sum_{p,q=N_\text{II}/2+1}^{N_\Omega} \Phi_q \Phi_p K_{pq} %\\
    = -\sideset{}{'}\sum_{p,q=1}^{N_\text{orb}} \phi_q \phi_p K_{pq} %\>\>\>\>\>\>\>\>\>\>\>\>\>\>\>\>\>\>\>\>\>\>\>\>\>\>\>\>\>\>\>\>\>\>\>\>\>\>\>\>\>\>
\end{multline}

where we have marked in blue the modifications with respect to GNOF. In particular, GNOFm adds static interpair interactions between occupied orbitals, and we have changed the constant from $\frac{1}{2}$ to $1$. The last equality holds only for real orbitals, as they fulfill $K_{pq} = L_{pq}$. 

\subsection{Basis Set Convergence Analysis}

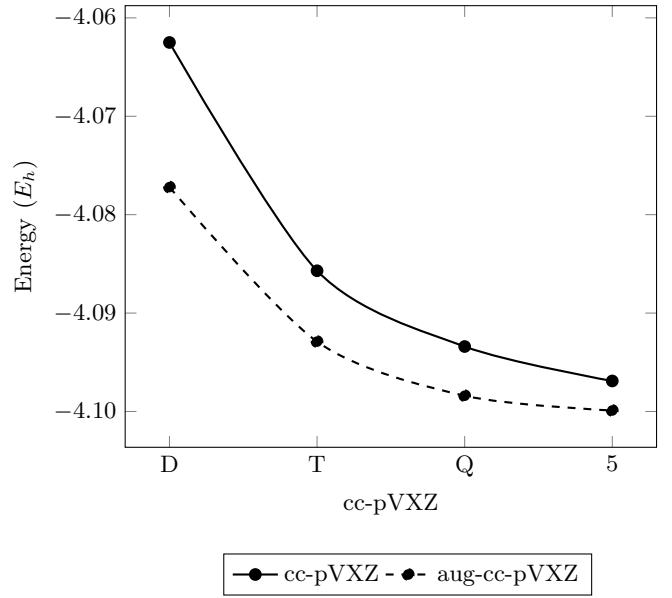
\begin{figure}[htb]
  \centering
  \begin{tikzpicture}
    \begin{axis}[
        xlabel=cc-pVXZ,
        ylabel= Energy ($E_h$),
        y tick label style={
            /pgf/number format/.cd,
            fixed,
            fixed zerofill,
            precision=2,
            /tikz/.cd
        },
        xticklabels={D,T,Q,5},
        xtick={2,...,5},
%        legend pos=north east,
        legend style={at={(0.65\linewidth,-0.22\linewidth)}},
        legend columns=2,
        width=\linewidth,
%        height=0.4\textwidth,
%        grid=major,
        scaled ticks=true,
    ]
    %\addplot[red,dotted,every mark/.append style={solid},mark=*,mark size=1pt] table[x=r, y=cvdz] {h2x2x2.dat};
    \addplot[black,thick,mark=*,smooth] coordinates {
        (2,-4.0625)(3,-4.0857)(4,-4.0934)(5,-4.0969)
    };
    \addlegendentry{cc-pVXZ};
    \addplot[black,dashed,thick,mark=*,smooth] coordinates {
        (2,-4.0772)(3,-4.0929)(4,-4.0984)(5,-4.0999)
    };
    \addlegendentry{aug-cc-pVXZ};
    \end{axis}
\end{tikzpicture}
  \caption{Total energy of the \ce{H8} cube at 2.5 \AA\ internuclear distance as a function of the cardinal number X in the cc-pVXZ basis set}
  \label{fig:S1}
\end{figure}

To evaluate the convergence of our results with respect to the size of the basis set and to assess the stability of the new NO optimization scheme, we performed calculations for the \ce{H8} cube using Dunning’s cc-pVXZ and aug-cc-pVXZ basis sets.\cite{BSE2019} This systematic approach allows us to analyze the impact of increasing basis set size on the computed results and to verify the robustness of our optimization method under these conditions. Fig.~\ref{fig:S1} presents the total energy as a function of X, where X indicates the basis set. The smooth behavior of the energy as X increases confirms the expected convergence trend.

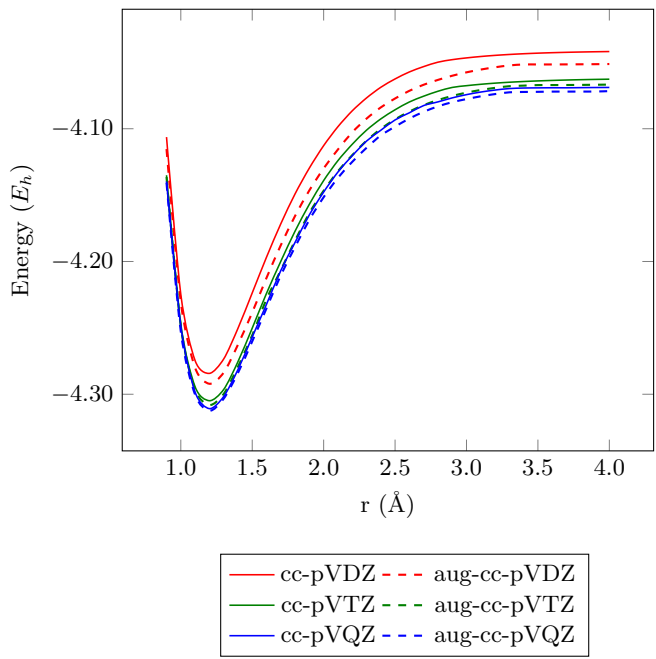
\begin{figure}[htb]
  \centering
  \begin{tikzpicture}
    \begin{axis}[
        xlabel=r (\AA),
        ylabel= Energy ($E_h$),
        x tick label style={
            /pgf/number format/.cd,
            fixed,
            fixed zerofill,
            precision=1,
            /tikz/.cd
        },
        y tick label style={
            /pgf/number format/.cd,
            fixed,
            fixed zerofill,
            precision=2,
            /tikz/.cd
        },
        %legend pos=south east,
        legend style={at={(0.65\linewidth,-0.22\linewidth)}},
        legend columns=2,
        ymax=-4.01,
        width=\linewidth,
%        height=0.4\textwidth,
%        grid=major,
        scaled ticks=true,
    ]
    %\addplot[red,dotted,every mark/.append style={solid},mark=*,mark size=1pt] table[x=r, y=cvdz] {h2x2x2.dat};
    \addplot[red, semithick, smooth] table[x=r, y=cvdz] {h2x2x2.dat};
    \addlegendentry{cc-pVDZ}
    \addplot[red, dashed, thick, smooth] table[x=r, y=acvdz] {h2x2x2.dat};
    \addlegendentry{aug-cc-pVDZ}
    \addplot[Green, semithick, smooth] table[x=r, y=cvtz] {h2x2x2.dat};
    \addlegendentry{cc-pVTZ}
    \addplot[Green, dashed, thick, smooth] table[x=r, y=acvtz] {h2x2x2.dat};
    \addlegendentry{aug-cc-pVTZ}
    \addplot[blue, semithick, smooth] table[x=r, y=cvqz] {h2x2x2.dat};
    \addlegendentry{cc-pVQZ}
    \addplot[blue, dashed, thick, smooth] table[x=r, y=acvqz] {h2x2x2.dat};
    \addlegendentry{aug-cc-pVQZ}
    \end{axis}
\end{tikzpicture}
  \caption{Dissociation curves of the \ce{H8} cube computed with different basis sets.}
  \label{fig:S2}
\end{figure}

Fig.~\ref{fig:S2} shows the dissociation curves for the \ce{H8} cube computed using the cc-pVXZ and aug-cc-pVXZ basis sets for X=\{D,T,Q\}. The overall agreement between the different basis sets demonstrates the stability of the NO optimization scheme and confirms that the dissociation trends remain consistent regardless of the choice of basis set. These results confirm that our approach remains stable even when we employ larger and more flexible basis sets.

\providecommand{\noopsort}[1]{}\providecommand{\singleletter}[1]{#1}%

\end{document}